\documentclass[prb,10pt,twocolumn]{revtex4}

\usepackage{epsfig}
\graphicspath{{/}}
\usepackage{graphicx}

\begin{document}

\title{A Many-Body Theory of the Optical Conductivity of Excitons and Trions in Two-Dimensional Materials}

\author{Farhan Rana, Okan Koksal, Christina Manolatou}
\affiliation{School of Electrical and Computer Engineering, Cornell University, Ithaca, NY 14853}
\email{fr37@cornell.edu}

\begin{abstract}
The optical spectra of two dimensional (2D) materials exhibit sharp absorption peaks that are commonly identified with exciton and trions (or charged excitons). In this paper, we show that excitons and trions in doped 2D materials can be described by two coupled Schr{\"o}dinger-like equations - one two-body equation for excitons and another four-body equation for trions. In electron-doped 2D materials, a bound trion state is identified with a four-body bound state of an exciton and an excited conduction band electron-hole pair. In doped 2D materials, the exciton and the trions states are the not the eigenstates of the full Hamiltonian and their respective Schr{\"o}dinger equations are coupled due to Coulomb interactions. The strength of this coupling increases with the doping density. Solutions of these two coupled equations can quantitatively explain all the prominent features experimentally observed in the optical absorption spectra of 2D materials including the observation of two prominent absorption peaks and the variation of their energy splittings and spectral shapes and strengths with the electron density. The optical conductivity obtained in our work satisfies the optical conductivity sum rule exactly. A superposition of exciton and trion states can be used to construct a solution of the two coupled Schr{\"o}dinger equations and this solution resembles the variational exciton-polaron state~\cite{Imam16}, thereby establishing the relationship between our approach and Fermi polaron physics~\cite{Demler12,Demler18,Chevy06}. 
\end{abstract}  
                                    
\maketitle

\section{Introduction}
Optical absorption and emission spectra of two-dimensional (2D) materials, most notably transition metal dichalcogenides (TMDs), exhibit distinct peaks that are attributed to neutral and charged excitons (or trions)~\cite{Fai13, Changjian14, Berk13, Chernikov14, Chernikov15}. Trions have been discussed extensively in the literature ~\cite{Changjian14,Imam16,Combes03,Combes12,Macdonald17,Suris01,Urba17}. In electron-doped materials, a trion state has been described in many different ways, i) as a bound state of two conduction band (CB) electrons and a valence band (VB) hole, or an electron bound to an exciton~\cite{Combes03,Combes12,Berk13}, ii) as a bound state of two CB electrons and a VB valence band hole, plus a CB hole~\cite{Changjian14,Chang19,Suris03}, iii) as an attractive exciton-polaron~\cite{Macdonald17}, and iv) as a bound molecular state similar to the one that appears in the literature on Fermi polaron physics~\cite{Imam16,Demler18}. The relationship between these different pictures is not clear. Sidler et al. considered attractive exciton-polarons to be different from trions and claimed to have seen optical signatures of both in experiments~\cite{Imam16}. Efimkin et al. identified the lower (higher) energy peak appearing in the absorption spectra of 2D materials (TMDs in particular) with attractive (repuslive) exciton-polarons. Earlier, Suris had explained these two absorption peaks in the context of 2D quantum well physics as mixed exction-trion states~\cite{Suris01b}. The trion state considered by Combescot et al.~\cite{Combes03}, which consisted of a bound state of two CB electrons and a VB hole and resembled a bound molecular state that appears in the literature on Fermi Polarons~\cite{Demler18}, where deemed to have negligible optical matrix element with the ground state by Sidler et al.~\cite{Imam16}. The variational trion state used previously by the authors~\cite{Changjian14}, which consisted of a bound state of two CB electrons and one VB hole, plus a CB hole, reproduced the measured trion optical absorption spectra in 2D transition metal dichalcogenides (TMDs) with fairly good accuracy in the low electron density limit but it could not explain the splitting of the trion and exciton absorption peaks as a function of the electron density, nor could it explain the transfer of the spectral weight in the optical absorption spectra from the exciton to the trion with the increase in the electron density. Diagrammatic perturbation theory involving summation of ladder diagrams corresponding to exciton-electron interactions as well as variational ansatz have been used to describe excitons interacting with electrons in electron-doped semiconductors~\cite{Imam16,Macdonald17,Chang19}. The solutions correspond to states that describe screening of the exciton by the electrons, or what are also called exciton-polarons. Exciton-polaron solutions have been successful in capturing the variations of the of the energy splittings as well as the spectral weight transfers observed in the optical absorption spectra as the electron density is varied. On the other hand, the three-body trion physics has been fairly successful in predicting the experimentally observed exciton-trion splittings (or the trion binding energies) in the limit of vanishing electron density ~\cite{Berk13, Changjian14,Urba17,Suris01}. The connection between all the above pictures of a trion remains unclear.

\begin{figure}
  \begin{center}
   \includegraphics[width=0.8\columnwidth]{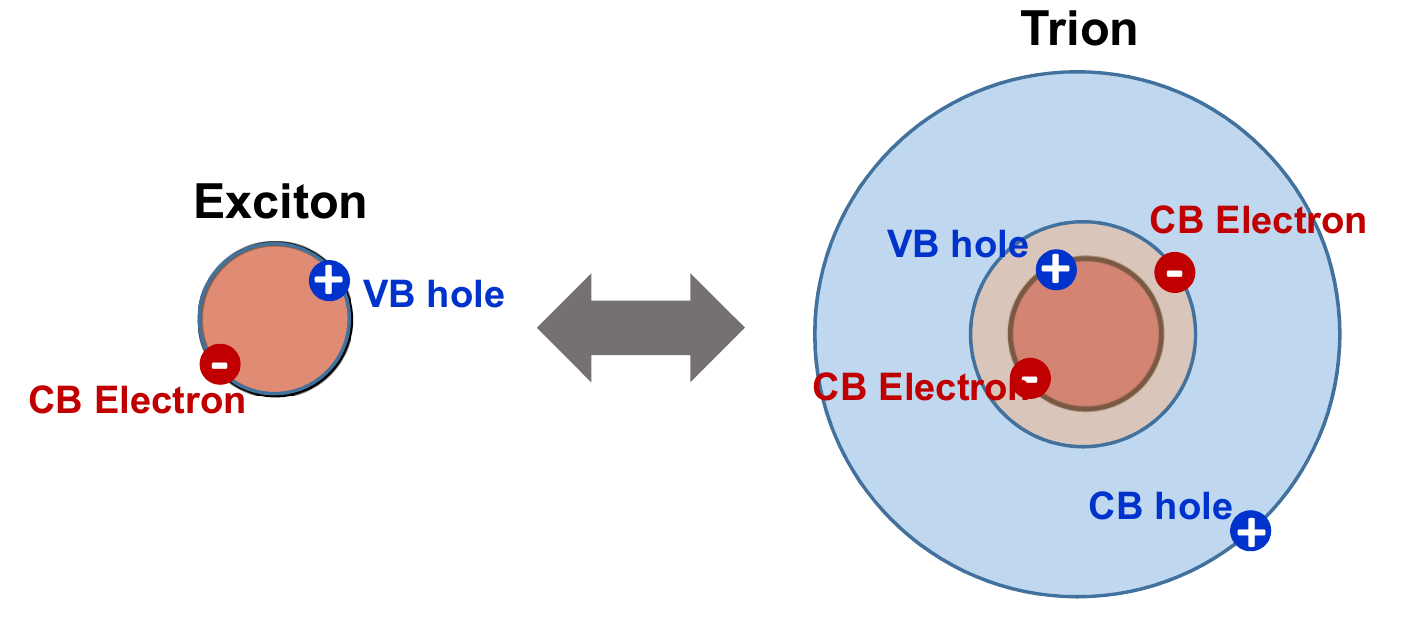}
    \caption{Bound exciton (left) and bound trion (right) states in an n-doped 2D material are depicted. A bound trion is a 4-body state and consists of two conduction band (CB) electrons, a valence band (VB) hole, and a conduction band (CB) hole. When the background electron density is non-zero, these two states are coupled due to electron-electron and electron-hole Coulomb interactions.}
    \label{fig:fig1}
  \end{center}
\end{figure}

In this paper, we use the many-body density matrix technique to describe describe excitons and trions in electron doped two-dimensional materials ~\cite{haugbook}. Our main results are summarized below.

{\em Excitons and trions in doped materials can be described by two coupled Schr{\"o}dinger-like equations}. One is the standard two-body Schr{\"o}dinger equation of a CB electron interacting with a VB hole (or an exciton)~\cite{haugbook}, and the other is a four-body Schr{\"o}dinger equation of two CB electrons interacting with a VB hole and a CB hole (See Fig.~\ref{fig:fig1}). A trimmed version of the latter, obtained by dropping the CB hole, is the standard three-body Schr{\"o}dinger equation that is commonly used to describe trions ~\cite{Changjian14, Berk13,Combes03,Combes12,Suris01,Urba17}. The CB hole is created when an electron is scattered out of the Fermi sea. Trions states are therefore identified with the eigenstates of this four-body Schr{\"o}dinger equation. These two Schr{\"o}dinger equations emerge naturally from the density matrix approach. The most interesting fact is that these two equations are coupled via Coulomb matrix elements that result from electron-electron and electron-hole interactions. Therefore, exciton and trion states are not independent excitations in doped materials. The solution of these two coupled equations provide a simple but exact description of an exciton interacting with CB electrons (within the purview of an exciton exciting only single electron-hole pair at a time in the conduction band). Solutions of these two coupled equations can quantitatively explain all the prominent features experimentally observed in the optical absorption spectra of 2D materials including the observation of two prominent absorption peaks and the variation of their energy splittings, spectral strengths, and spectral linewidths with the electron density. Interestingly, the exact solution of these two equations resembles the variational exciton-polaron solution,~\cite{Imam16,Chang19} thereby establishing the relationship between the two approaches. The solutions obtained in this work also explicitly satisfy the optical conductivity sum rule exactly at all carrier densities. 

The eigenspectrum of the four-body Schr{\"o}dinger equation has bound states, which are bound states of an exciton and a CB electron-hole pair, and unbound states, which are exciton-electron scattering states. However, these eigenstates of the four-body Schr{\"o}dinger equation as well as the eigenstates of the two-body Schr{\"o}dinger equation are not eigenstates of the full Hamiltonian in the presence of a non-zero electron density because of Coulomb interactions. The eigenstates of the full Hamiltonian are more accurately superpositions of two-body exciton and four-body trion states. The two prominent peaks observed in the optical absorption spectra of doped 2D materials correspond to these superpositon eigenstates. The essential physics is as follows. An exciton can Coulomb excite a CB electron-hole pair and bind it to form a trion. But it is  energetically favorable for the exciton to release this bound CB electron-hole pair and excite another CB electcron-hole pair and bind it (or scatter from it). These virtual transitions can occur at the lowest electron densities, even when the Fermi energy is much smaller than the binding energy of the four-body trion state. The energy thus gained is responsible for the observed near-linear increase in the energy splitting of the two absorption peaks with the Fermi energy~\cite{Fai13,Chernikov15}. The picture described here also captures the essential physics associated with the exciton-polarons. The main two peaks observed experimentally in the optical absorption spectra result from the coupling between the two-body exciton and four-body trion states and have also been identified as the repulsive and attractive exciton-polarons absorption peaks by Efimkin et al.~\cite{Macdonald17}. A complete and accurate description of the problem requires using both the bound and the unbound states of the four-body Schr{\"o}dinger equation. Exciton-electron scattering is thereby also included in our analysis. 

In this paper, we have focused on single CB electron-hole pair excitation by an exciton and have ignored multiple electron-hole pair excitations that are expected to become important when the Fermi energy approaches the exciton binding energy~\cite{Macdonald17}. Effects related to Fermi edge singularities~\cite{Mahan1,Mahan2}, which involve multiple electron-hole pair excitations and are expected to become important if the hole mass or the exciton mass were much bigger than the electron mass (which is not the case in most 2D materials including TMDs), are therefore also ignored. Our work shows that a description based on single pair excitations can adequately explain the prominent features of the experimentally measured optical absorption spectra of 2D materials.  

In the light of the introductory discussion above, the name "trion'' seems like a misnomer since the involved states are either two-body or four-body states. But given the long history of the use of this terminology, in this paper we will use the term ``trion'' for the eigenstates of the four-body Schr{\"o}dinger equation (Fig.~\ref{fig:fig1}). After this work had been completed, we became aware of the earlier work by Suris~\cite{Suris03} in which mixed exciton-trion modes had been introduced to account for the coupling between the excitons and the trions resulting from Coulomb interactions. Suris was also the first one to argue that it would be energetically favorable for the CB hole to bind to the two CB electrons and one VB hole in a bound trion state as result of trion-exciton coupling. The work presented here is conceptually along the same lines, but our approach is different in many ways. The use of coupled two-body and four-body Schr{\"o}dinger equations enables one to include the effects of the electron density, as well as the effects of exciton-trion coupling, on the wavefunctions and the binding energies of excitons and trions. Our approach, which does not involve wavevector-independent Coulomb potentials and artificial cut-offs of momentum integrals~\cite{Imam16,Macdonald17,Suris01b}, enables us to obtain quantitative results for excitons and trions in two dimensional materials. Suris had also ignored direct Coulomb interactions between the CB hole and the three other particles in a bound trion state. Our four-body Schr{\"o}dinger equation includes these interactions which cannot be ignored at the moderate to high electron densities at which trion signatures are experimentally observed in 2D materials. In agreement with the reported experimental results on 2D materials, but not in agreement with the conclusions drawn by Suris, we do not find any broad absorption bands in the optical absorption spectra associated with trions states that do not have a bound CB hole. In contrast to other previous works~\cite{Imam16,Macdonald17,Chang19}, our approach also explicitly takes into account electron's spin and valley degrees of freedom and exchange interactions. Finally, our approach also sheds light on the debated question of whether or not trions have appreciable optical oscillator strength. Our results show that the trion states, defined in this paper as the eigenstates the four-body Schr{\"o}dinger equation, have no optical oscillator strength and their optical activity results only from their Coulomb coupling to the exciton states.

\section{Theoretical Model} \label{sec:thmodel}
In this Section we set up the Hamiltonian and derive the main equations. Although the focus is on electron-doped 2D TMD materials, the arguments are kept general enough to be applicable to any 2D material.    

\subsection{Hamiltonian} \label{subsec:hamiltonian}
We consider a 2D TMD monolayer at $z=0$. Light with a small in-plane momentum $\vec{Q}$ is incident on the layer. The Hamiltonian describing electrons and holes in the TMD layer (near the $K$ and $K'$ points in the Brillouin zone) interacting with each other and with the optical mode in the rotating wave approximation is~\cite{Xiao12,Changjian14,HWang16,Mano16},
\begin{eqnarray}
H & = & \sum_{\vec{k},s} E_{c,s}(\vec{k}) c_{s}^{\dagger}(\vec{k})c_{s}(\vec{k}) + \sum_{\vec{k},s} E_{v,s}(\vec{k}) b_{s}^{\dagger}(\vec{k})b_{s}(\vec{k}) \nonumber \\
& + & \frac{1}{A}\sum_{\vec{q},\vec{k},\vec{k}',s,s'} U(q)  c_{s}^{\dagger}(\vec{k}+\vec{q})b_{s'}^{\dagger}(\vec{k}'-\vec{q})b_{s'}(\vec{k}')c_{s}(\vec{k}) \nonumber \\
& + & \frac{1}{2A}\sum_{\vec{q},\vec{k},\vec{k}',s,s'} V(q)  c_{s}^{\dagger}(\vec{k}+\vec{q})c_{s'}^{\dagger}(\vec{k}'-\vec{q})c_{s'}(\vec{k}')c_{s}(\vec{k}) \nonumber \\
& + & \hbar \omega(\vec{Q}) a^{\dagger}(\vec{Q})a(\vec{Q}) \nonumber \\
& + & \frac{1}{\sqrt{A}}\sum_{\vec{k,s}} \left( g_{s}c_{s}^{\dagger}(\vec{k}+\vec{Q})b_{s}(\vec{k})a(\vec{Q})  + h.c \right)
\label{eq:H}
\end{eqnarray}
Here, $E_{c,s}(\vec{k})$ and $E_{v,s}(\vec{k})$ are the conduction and valence band energies. $s,s'$ represent the spin/valley  degrees of freedom in the 2D material, and we assume for simplicity that the electron and hole effective masses are independent of the spin/valley. $U(\vec{q})$  represents Coulomb interaction between electrons in the conduction and valence bands and $V(\vec{q})$ represents Coulomb interaction among the electrons in the conduction bands. $\hbar \omega(\vec{Q})$ is the energy of a photon with in-plane momentum $\vec{Q}$, and $g_{s}$ is the electron-photon coupling constant. Other than for phase factors that are not relevant in the discussion that follows, $g_{s}$ for electron states near the band edges in 2D TMDs can be given by~\cite{HWang16,Mano16}, 
\begin{equation}
  g_{s} = ev\sqrt{\frac{\hbar}{2\langle \epsilon \rangle\omega(\vec{Q})}} \, \chi(z=0)
\end{equation} 
where, $v$ is the interband velocity matrix element~\cite{Xiao12,Changjian14,HWang16,Mano16}, and $\chi (z)$ describes the amplitude of the optical mode in the z-direction.    

\subsection{Density Matrix Approach} \label{sec:eom}
We use a many body density matrix approach which has been fairly successful in modeling exciton physics~\cite{haugbook,Kira12}, and it has also been previously used for trions in the limit of vanishingly small electron densities~\cite{Esser01}.

We start from the Heisenberg equation for the photon operator, which after averaging, is~\cite{haugbook,Kira12},
\begin{equation}
 \left[ \hbar \omega(\vec{Q}) +  i\hbar \frac{\partial}{\partial t} \right] \langle a^{\dagger}(\vec{Q},t) \rangle =  - \frac{1}{\sqrt{A}} \sum_{\vec{k},s} g_{s}P_{\vec{Q}}(\vec{k},s;t)  
  \end{equation}
The polarization $P_{\vec{Q}}(\vec{k},s;t)$ equals the equal-time two-body correlation $\langle c_{s}^{\dagger}(\vec{k}+\vec{Q},t)b_{s}(\vec{k},t) \rangle$.  Assuming from now onwards that in steady state all the relevant equal-time correlation functions have the time dependence $e^{\pm i \omega t}$, the above equation becomes,
\begin{equation}
  \left[ \hbar \omega(\vec{Q}) + i\epsilon  - \hbar \omega \right] \langle a^{\dagger}(\vec{Q}) \rangle  =  - \frac{1}{\sqrt{A}} \sum_{\vec{k},s} g_{s}P_{\vec{Q}}(\vec{k},s) \label{eq:photon}
\end{equation}
where $P_{\vec{Q}}(\vec{k},s;t) = P_{\vec{Q}}(\vec{k},s)e^{\- i \omega t}$. The equation for $P_{\vec{Q}}(\vec{k},s)$ is,
\begin{eqnarray}
  & &  \left[ E_{c,s}(\vec{k}+\vec{Q}) - E_{v,s}(\vec{k}) + i\gamma_{ex} - \hbar \omega \right] P_{\vec{Q}}(\vec{k},s) = \nonumber \\
  & & - \frac{1}{\sqrt{A}} g^{*}_{s} \langle a^{\dagger}(\vec{Q}) \rangle \left[ 1 - f_{c,s}(\vec{k}+\vec{Q}) \right] \nonumber \\
  & & + \frac{1}{A}\sum_{\vec{q}} U(\vec{q}) P_{\vec{Q}}(\vec{k} + \vec{q},s)  \left[ 1 - f_{c,s}(\vec{k} + \vec{Q}) \right] \nonumber \\
  & & -\frac{1}{A} \sum_{\vec{q},\vec{p},s'} U(\vec{q}) \nonumber \\
  & & \times T^{c}_{\vec{Q}}(\vec{k} + (\xi + \eta)\vec{Q} - \xi \vec{p},s;(\xi + \eta)\vec{p} - \xi \vec{Q} - \vec{q},s';\vec{p},s') \nonumber \\
  & & +\frac{1}{A} \sum_{\vec{q},\vec{p},s'} V(\vec{q}) \nonumber \\
  & & \times T^{c}_{\vec{Q}}(\vec{k} + (\xi + \eta)\vec{Q} - \xi \vec{p} +\vec{q},s;(\xi + \eta)\vec{p} - \xi \vec{Q} - \vec{q},s';\vec{p},s') \nonumber \\
  \label{eq:exciton1}
  \end{eqnarray}
Here, $\langle c_{s}^{\dagger}(\vec{k}) c_{s}(\vec{k}) \rangle = f_{c,s}(\vec{k})$ is the electron occupation probability, $\gamma_{ex}$ is a phenomenological decoherence rate for the polarization that includes dephasing due to all processes other than exciton-electron scattering. The energies $E_{c,s}(\vec{k})$ include renormalizations due to exchange at the Hartree-Fock level ($-(1/A)\sum_{\vec{q}} V(\vec{q}) f_{c,s}(\vec{k}-\vec{q})$). $\lambda_{h} = 1-\lambda_{e} = m_{h}/m_{ex}$  ($m_{ex} = m_{e} + m_{h}$), where $m_{e}$ ($m_{h}$) is the electron (hole) effective mass. (\ref{eq:exciton1}), without the first term and the last two terms on the RHS, is the standard eigenvalue equation for excitons~\cite{haugbook,Kira12}. (\ref{eq:exciton1}) is not Hermitian but it can be converted into a Hermitian equation by rescaling the eigenfunctions (see Appendix~\ref{app1}). The Hermitian equation has eigenvalues $E^{ex}_{n}(\vec{Q},s)$ and eigenfunctions $\phi^{ex}_{n,\vec{Q}}(\vec{k}+\lambda_{h}\vec{Q})$. The eigenfunctions form a complete set. 

\begin{figure}
  \begin{center}
   \includegraphics[width=0.6\columnwidth]{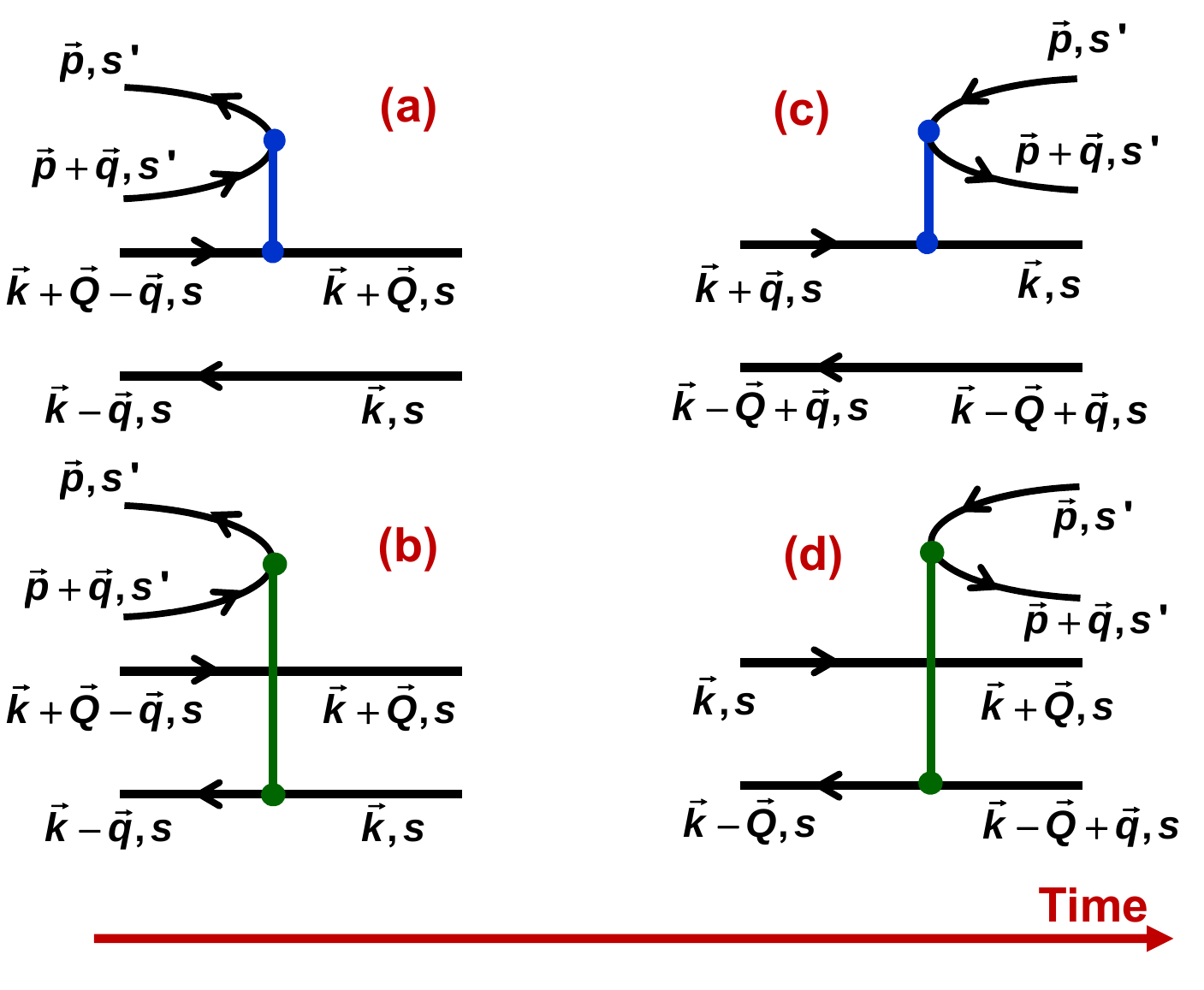}
   \caption{A sampling of diagrams and processes that connect two-body and four-body correlations in (\ref{eq:exciton1}) and (\ref{eq:4body1}). Coulomb interactions are depicted by the vertical lines. The horizontal lines are electron and hole propagators. Diagrams corresponding to exchange processes are not shown.}
    \label{fig:vertex}
  \end{center}
\end{figure}

The last two terms in (\ref{eq:exciton1}) on the RHS contain four-body correlations $T_{\vec{Q}}$ and correspond to the diagrams shown in Fig.~\ref{fig:vertex}(a,b). Assuming that $m_{tr} = 2m_{e} + m_{h}$, $\xi = m_{e}/m_{tr}$, and $\eta = m_{h}/m_{tr}$, we define a four-body equal-time correlation $T_{\vec{Q}}(\vec{k}_{1},s_{1};\vec{k}_{2},s_{2};\vec{p},s_{2};t)$ as follows,
\begin{equation}
\langle c_{s_{1}}^{\dagger}(\vec{\underline{k}}_{1},t) c_{s_{2}}^{\dagger}(\vec{\underline{k}}_{2},t) b_{s_{1}}(\vec{\underline{k}}_{1}+\vec{\underline{k}}_{2} - (\vec{Q}+\vec{p}),t) c_{s_{2}}(\vec{p},t) \rangle
\end{equation}
The underlined vector $\vec{\underline{k}}$ stands for $\vec{k}+\xi (\vec{Q}+\vec{p})$. $T_{\vec{Q}}$ describes the correlations arising from Coulomb interactions among four particles: two CB electrons, a VB hole, and a CB hole. $\vec{Q}$ is the total momentum of this 4-body state. We also define a {\em fully connected} four-body correlation (as defined in the cluster expansion technique~\cite{Kira12}),
\begin{eqnarray}
  & &  T_{\vec{Q}}(\vec{k}_{1},s_{1};\vec{k}_{2},s_{2};\vec{p},s_{2};t) = T^{c}_{\vec{Q}}(\vec{k}_{1},s_{1};\vec{k}_{2},s_{2};\vec{p},s_{2};t) \nonumber \\
  & & - f_{c,s_{2}}(\vec{p}) P_{\vec{Q}}(\vec{\underline{k}}_{1} -\vec{Q},s_{1}) \delta_{\vec{\underline{k}}_{2},\vec{p}} \nonumber \\
  & & + f_{c,s_{2}}(\vec{p}) P_{\vec{Q}}(\vec{\underline{k}}_{2} - \vec{Q}) \delta_{s_{1},s_{2}} \delta_{\vec{\underline{k}}_{2},\vec{p}} 
\end{eqnarray}

The equation for the connected correlation $T^{c}_{\vec{Q}}(\vec{k}_{1},s_{1};\vec{k}_{2},s_{2};\vec{p},s_{2})$ is found to be,
\begin{eqnarray}
  & & \left[ E_{c,s_{1}}(\vec{\underline{k}}_{1}) + E_{c,s_{2}}(\vec{\underline{k}}_{2}) - E_{v,s_{1}}(\vec{\underline{k}}_{1}+\vec{\underline{k}}_{2} - (\vec{Q}+\vec{p})) \right. \nonumber \\
  & &  \left. - E_{c,s_{2}}(\vec{p}) + i\gamma_{tr} - \hbar \omega \right] T^{c}_{\vec{Q}}(\vec{k}_{1},s_{1};\vec{k}_{2},s_{2};\vec{p},s_{2}) = \nonumber \\
  & & -\frac{1}{A} \sum_{\vec{q}} V(\vec{q}) T^{c}_{\vec{Q}}(\vec{k}_{1}+\vec{q},s_{1};\vec{k}_{2}-\vec{q},s_{2};\vec{p},s_{2}) \nonumber \\
  & & \times \left[ 1-f_{c,s_{1}}(\vec{\underline{k}}_{1})  - f_{c,s_{2}}(\vec{\underline{k}}_{2}) \right] \nonumber \\
  & & +\frac{1}{A} \sum_{\vec{q}} U(\vec{q}) T^{c}_{\vec{Q}}(\vec{k}_{1}+\vec{q},s_{1};\vec{k}_{2},s_{2};\vec{p},s_{2}) \left[ 1-f_{c,s_{1}}(\vec{\underline{k}}_{1})  \right] \nonumber \\
  & & +\frac{1}{A} \sum_{\vec{q}} U(\vec{q}) T^{c}_{\vec{Q}}(\vec{k}_{1},s_{1};\vec{k}_{2}-\vec{q},s_{2};\vec{p},s_{2}) \left[ 1-f_{c,s_{2}}(\vec{\underline{k}}_{2}) \right] \nonumber \\
  & & +\frac{1}{A} \sum_{\vec{q}} V(\vec{q}) T^{c}_{\vec{Q}}(\vec{k}_{1}+ (\xi + \eta)\vec{q},s_{1};\vec{k}_{2}-\xi \vec{q},s_{2};\vec{p}+\vec{q},s_{2}) \nonumber \\
  & & \times \left[f_{c,s_{2}}(\vec{p}) - f_{c,s_{1}}(\vec{\underline{k}}_{1}) \right] \nonumber \\
  & & +\frac{1}{A} \sum_{\vec{q}} V(\vec{q}) T^{c}_{\vec{Q}}(\vec{k}_{1} - \xi\vec{q},s_{1};\vec{k}_{2} + (\xi +\eta) \vec{q},s_{2};\vec{p}+\vec{q},s_{2}) \nonumber \\
  & & \times \left[f_{c,s_{2}}(\vec{p}) - f_{c,s_{2}}(\vec{\underline{k}}_{2})  \right] \nonumber \\
  & & -\frac{1}{A} \sum_{\vec{q}} U(\vec{q}) T^{c}_{\vec{Q}}(\vec{k}_{1}-\xi \vec{q},s_{1};\vec{k}_{2}- \xi \vec{q},s_{2};\vec{p}+\vec{q},s_{2}) f_{c,s_{2}}(\vec{p}) \nonumber \\
  & & + \frac{f_{c,s_{2}}(\vec{p})}{A} \sum_{\vec{q}} V(\vec{q}) \left[ 1-f_{c,s_{1}}(\vec{\underline{k}}_{1}) - f_{c,s_{2}}(\vec{\underline{k}}_{2})  \right] \nonumber \\
  & & \times \left[ P_{\vec{Q}}(\vec{\underline{k}}_{1} - \vec{Q} + \vec{q},s_{1}) \delta_{\vec{\underline{k}}_{2}-\vec{q},\vec{p}} \right. \nonumber \\
  & & \left. - P_{\vec{Q}}(\vec{\underline{k}}_{2} - \vec{Q} - \vec{q},s_{2}) \delta_{\vec{\underline{k}}_{1}+\vec{q},\vec{p}} \delta_{s_{1},s_{2}} \right] \nonumber \\
  & & - \frac{f_{c,s_{2}}(\vec{p})}{A} \sum_{\vec{q}} U(\vec{q}) \left\{ P_{\vec{Q}}(\vec{\underline{k}}_{1} - \vec{Q},s_{1}) \delta_{\vec{\underline{k}}_{2}-\vec{q},\vec{p}} \left[ 1-f_{c,s_{2}}(\vec{\underline{k}}_{2})  \right] \right. \nonumber \\
  & & - \left. P_{\vec{Q}}(\vec{\underline{k}}_{2} - \vec{Q},s_{2}) \delta_{\vec{\underline{k}}_{1}+\vec{q},\vec{p}} \delta_{s_{1},s_{2}} \left[ 1-f_{c,s_{1}}(\vec{\underline{k}}_{1})  \right] \right\} \nonumber \\
  \label{eq:4body1}
\end{eqnarray}
In deriving the above equation, all six-body correlations are reduced to four-body correlations using the cluster expansion~\cite{Kira12}. By ignoring higher order correlations we are ignoring the generation of multiple particle-hole pairs in the CB. Here, $\gamma_{tr}$ is a phenomenological decoherence rate. If $\vec{r}_{e1}$, $\vec{r}_{e2}$, $\vec{r}_{h1}$, are $\vec{r}_{h2}$ the coordinates of the two electrons, the VB hole, and the CB hole, respectively, then $\vec{k}_{1}$, $\vec{k}_{2}$, $\vec{Q}$, and $\vec{p}$ are the momenta associated with the coordinates $\vec{r}_{e1} - \vec{r}_{h1}$, $\vec{r}_{e2} - \vec{r}_{h1}$, $\vec{R} = \xi (\vec{r}_{e1} + \vec{r}_{e2}) + \eta \vec{r}_{h1}$, and $\vec{R} - \vec{r}_{h2}$, respectively. Here, $\vec{R}$ is the center of mass coordinate of the two electrons and the VB hole. Ignoring the last two terms on the RHS in (\ref{eq:4body1}) that involve $P_{\vec{Q}}$, Fourier transform of the remaining terms will result in a four-body Schr{\"o}dinger equation. Each term on the RHS in the above equation (except the last two) describes Coulomb interaction between two of the four particles.The last two terms involving $P_{\vec{Q}}$ capture the generation of four-body correlation from two-body correlation, or the creation of an CB electron-hole pair by an exciton, and and correspond to the diagrams shown in Fig.~\ref{fig:vertex}(c,d). We should mention here that an equation similar to (\ref{eq:4body1}) was obtained by Esser et al.~\cite{Esser01}, but in that work the connected nature of $T^{c}_{\vec{Q}}$ was overlooked, the terms containing interactions with the CB hole were ignored, the phase-space restricting factors were ignored too, and, most importantly, the terms containing the polarization $P_{\vec{Q}}$ were also missed. Ignoring the coupling to $P_{\vec{Q}}$ in (\ref{eq:4body1}) is equivalent to ignoring exciton-trion coupling via Coulomb interactions.

(\ref{eq:exciton1}) and (\ref{eq:4body1}) are a closed system of coupled Schr{\"o}dinger equations for two-body and four-body systems.

\subsection{Trion States} \label{sec:trions}
The trion states are defined here as the eigenstates of the four-body Schr{\"o}dinger equation given in (\ref{eq:4body1}). (\ref{eq:4body1}) is not Hermitian but it can be converted into a Hermitian equation (see Appendix~\ref{app2}), with a few suitable approximations, and the eigenfunctions therefore form a complete set. The eigenfunctions are written as $\phi^{tr}_{m,\vec{Q}}(\vec{k}_{1},s_{1};\vec{k}_{2},s_{2};\vec{p},s_{2})$. The corresponding eigenenergies are $E^{tr}_{m}(\vec{Q},s_{1},s_{2})$.  The eigenstates include bound four-body states, unbound exciton-electron scattering states, and completely unbound four-body states. The latter have high energies and may be ignored here. The eigenfunctions $\phi^{tr}_{m,\vec{Q}}(\vec{k}_{1},s_{1};\vec{k}_{2},s_{2};\vec{p},s_{2})$ are either symmetric or antisymmetric in $\vec{k}_{1}$ and $\vec{k}_{2}$ depending on the values of $s_{1}$ and $s_{2}$ and on the spin state of the two electrons (singlet or triplet).  

\subsection{The Fate of the Conduction Band Hole} \label{sec:hole}
The 4th, 5th, and 6th terms on the RHS in the four-body Schr{\"o}dinger equation include interactions involving the CB hole that is generated when an electron is scattered out of the Fermi sea by the exciton to form a trion. The CB hole wavefunction can have a radius no smaller than $\sim 1/k_{F}$ in real space and therefore it can be much larger than the exciton and the three-body trion radii at small electron densities. For this reason, interactions involving the CB hole have been ignored in previous works ~\cite{Imam16,Macdonald17,Chang19,Suris03}. Here we argue that the CB hole needs to be taken into account in bound trion states. Signatures of bound trion states are observed only at moderate to high electron densities in 2D materials ($n > 10^{12}$ cm$^{-2}$) at which Fermi energy can be appreciable. The LHS of (\ref{eq:4body1}) has the energies of the photoexcited CB electron, the VB hole, and the initial and final energies of the CB electron scattered out of the Fermi sea. All energies include renormalization due to exchange. An electron within the Fermi sea has a larger energy renormalization than an electron well outside the Fermi sea~\cite{haugbook}. Consequently, when an electron is scattered out of the Fermi sea to bind to an exciton and form a tightly bound trion state, the difference in its initial and final exchange energies, as given by the terms on the LHS of (\ref{eq:4body1}), can be pretty large - so much so that a bound trion state may not even be energetically possible except at very small electron densities. The inclusion of the terms on the RHS of (\ref{eq:4body1}), which include Coulomb interactions involving the CB hole, make up for this energy difference and make bound trion states possible and energetically favorable provided the CB hole, along with the CB electron, gets bounded to the exciton to make a four-body bound state depicted in Fig.(\ref{fig:fig1}). In addition to direct Coulomb interactions involving the CB hole, and as shown in the earlier work of Suris~\cite{Suris03}, exciton-trion coupling, and the energy gained therewith, also favors the binding of the CB hole bound to the exciton and the CB electron in a four-body bound state.          

\subsection{Exciton Self-Energy and Optical Conductivity}

A formal solution of (\ref{eq:4body1}) can be written in terms of its eigenfunctions as,
\begin{eqnarray}
  & & T^{c}_{\vec{Q}}(\vec{k}_{1},s_{1};\vec{k}_{2},s_{2};\vec{p},s_{2}) = - (1 + \delta_{s_{1},s_{2}}) \nonumber \\
  & & \times \sum_{m} \sqrt{f_{c,s_{2}}(\vec{p}) \left[ 1-f_{c,s_{1}}(\vec{\underline{k}}_{1}) \right]\left[ 1 - f_{c,s_{2}}(\vec{\underline{k}}_{2})  \right]} \nonumber \\
  && \times \frac{\phi^{tr}_{m,\vec{Q}}(\vec{k}_{1},s_{1};\vec{k}_{2},s_{2};\vec{p},s_{2})}{\hbar\omega - E^{tr}_{m}(\vec{Q},s_{1},s_{2}) - i\gamma_{tr}} \nonumber \\
  & & \frac{1}{A^{4}} \sum_{\vec{k}'_{1},\vec{p}',\vec{q}} \phi^{tr*}_{m,\vec{Q}}(\vec{k}'_{1},s_{1};(\xi + \eta)\vec{p}'-\xi \vec{Q} + \vec{q},s_{2};\vec{p}',s_{2}) \nonumber \\
& & \times  \sqrt{ f_{c,s_{2}}(\vec{p}') \left[ 1 - f_{c,s_{2}}(\vec{p}'+\vec{q}) \right]} \left\{  \frac{\sqrt{1-f_{c,s_{1}}(\vec{\underline{k}}'_{1}) }}{}  \right. \nonumber \\
  & & \times   \left. V(\vec{q}) P_{\vec{Q}}(\vec{\underline{k}}'_{1} - \vec{Q} + \vec{q},s_{1})   -  U(\vec{q}) \frac{P_{\vec{Q}}(\vec{\underline{k}}'_{1} - \vec{Q},s_{1})}{\sqrt{1-f_{c,s_{1}}(\vec{\underline{k}}'_{1})}}    \right\} \nonumber \\
\end{eqnarray}
The summation over $m$ above implies a summation over all bound and unbound trion states consistent with the values of $s_{1}$ and $s_{2}$. The above solution can be used in (\ref{eq:exciton1}) to obtain the polarization,
\begin{eqnarray}
  & &  P_{\vec{Q}}(\vec{k},s) = \frac{g^{*}_{s}}{\sqrt{A}}  \langle a^{\dagger}(\vec{Q}) \rangle \sum_{n} \sqrt{1-f_{c,s}(\vec{k} + \vec{Q})} \nonumber \\
  & & \times \frac{\phi^{ex}_{n,\vec{Q}}(\vec{k} + \lambda_{h}\vec{Q})}{\hbar\omega - E^{ex}_{n}(\vec{Q},s) - i\gamma_{ex} - \Sigma^{ex*}_{n,s}(\vec{Q},\omega)} \nonumber \\
  & & \int \frac{d^{2}\vec{k}'}{(2\pi)^{2}} \phi^{ex*}_{n,\vec{Q}}(\vec{k}' + \lambda_{h}\vec{Q}) \sqrt{1-f_{c,s}(\vec{k}' + \vec{Q})} \nonumber \\
  \end{eqnarray}
Here, the summation over $n$ implies a summation over all bound and unbound exciton states. The expression for the exciton self energy $\Sigma^{ex}_{n,s}(\vec{Q},\omega)$ is given below. The above result can be used in (\ref{eq:photon}) to obtain an expression for the optical conductivity $\sigma(\vec{Q},\omega)$ of the 2D material,
\begin{eqnarray}
  &&  \sigma(\vec{Q},\omega) = i\frac{e^{2}v^{2}}{\omega} \nonumber \\
  && \times \sum_{n,s} \frac{\displaystyle \left|\int \frac{d^{2}\vec{k}'}{(2\pi)^{2}} \phi^{ex}_{n,\vec{Q}}(\vec{k}' + \lambda_{h}\vec{Q}) \sqrt{1-f_{c,s}(\vec{k}' + \vec{Q})} \right|^{2}}{\hbar\omega - E^{ex}_{n}(\vec{Q},s) + i\gamma_{ex} - \Sigma^{ex}_{n,s}(\vec{Q},\omega)} \nonumber \\
  \label{eq:cond1}
  \end{eqnarray}
The exciton self-energy can be expressed as,
\begin{eqnarray}
  &&  \Sigma^{ex}_{n,s}(\vec{Q},\omega) = \sum_{m,s'} \frac{ (1 + \delta_{s,s'}) \left| M_{m,n}(\vec{Q},s,s') \right|^{2}}{\hbar\omega - E^{tr}_{m}(\vec{Q},s,s') + i\gamma_{tr} }  \nonumber \\
  \label{eq:self1}
\end{eqnarray}
The summation over $m$ above implies a summation over all bound and unbound trion states consistent with the values of $s$ and $s'$. $M_{m,n}(\vec{Q},s,s')$ equals,
\begin{eqnarray}
& & \frac{1}{A^{3}} \sum_{\vec{k},\vec{p},\vec{q}} \phi^{tr*}_{m,\vec{Q}}(\vec{k}-\xi(\vec{p}+\vec{Q}),s;(\xi + \eta)\vec{p}-\xi \vec{Q} + \vec{q},s';\vec{p},s') \nonumber \\
  & & \times  \sqrt{ f_{c,s'}(\vec{p}) \left[ 1 - f_{c,s'}(\vec{p}+\vec{q}) \right]} \nonumber \\
  & & \times \left\{  \sqrt{\left[ 1-f_{c,s}(\vec{k}) \right]\left[ 1-f_{c,s}(\vec{k} + \vec{q}) \right] } \,\,\, V(\vec{q}) \right. \nonumber \\
  & & \times   \left. \phi^{ex}_{n,\vec{Q}}(\vec{k} - \lambda_{e} \vec{Q} + \vec{q})   -  U(\vec{q}) \phi^{ex}_{n,\vec{Q}}(\vec{k} - \lambda_{e} \vec{Q})   \right\} \nonumber \\
  \label{eq:M}
\end{eqnarray}
It is evident from the above expression that the coupling term $M_{m,n}(\vec{Q},s,s')$ increases with the electron density. The self-energy expression in (\ref{eq:self1}) assumes that an exciton state does not couple to a different exciton state due to Coulomb interactions via an intermediate trion state. Given that the energy separation between the lowest energy exciton state and the higher energy exciton states in 2D materials can be in the hundreds of milli electron volts range~\cite{Chernikov14}, the approximation made in this assumption is expected to be very good for the lowest energy exciton state.

\begin{figure}
  \begin{center}
   \includegraphics[width=0.9\columnwidth]{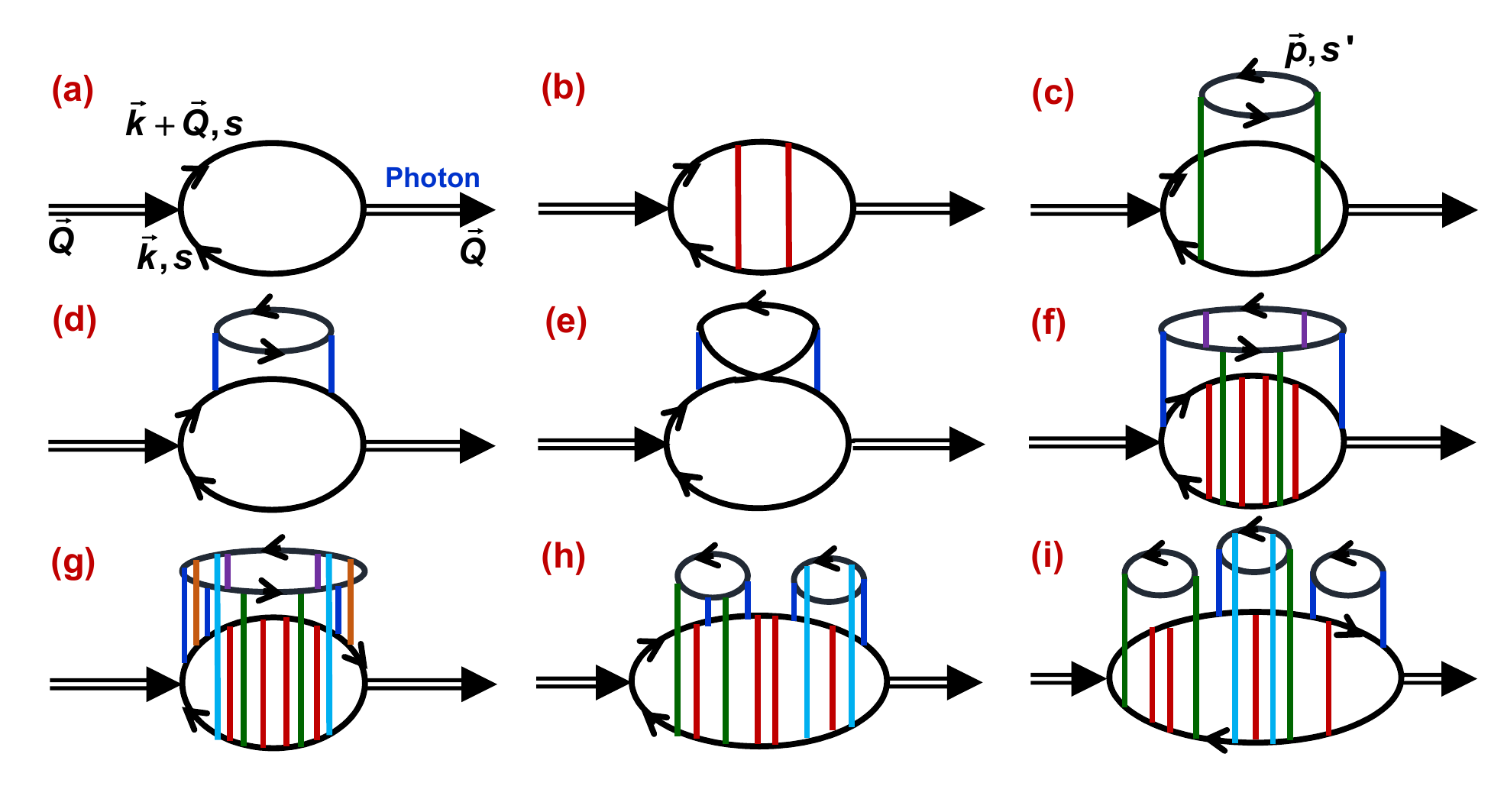}
   \caption{A sampling of diagrams and processes that contribute to the optical conductivity. Curved horizontal lines represent electron and hole propagators. The double horizontal line is the photon propagator. Coulomb interactions are depicted by the vertical lines. The small circles represent electron-hole pairs excited in the conduction band by the photogenerated exciton. Only fully connected diagrams contribute to the optical conductivity. Diagrams with multiple electron-hole pair excitations in the conduction band are not included in this work.}
    \label{fig:fig2}
  \end{center}
\end{figure}

The expressions in (\ref{eq:cond1}) and (\ref{eq:self1}) constitute the main results of this work. The optical conductivity in (\ref{eq:cond1}) corresponds to the diagrams and processes shown in Fig.~\ref{fig:fig2}. Only fully connected diagrams contribute to the optical conductivity. Diagrams with multiple particle-hole excitations at the same time are not included in this work. 

\subsection{Unbound Trion States: Exciton-Electron Scattering} \label{sec:ees}
The expression for the exciton self-energy in (\ref{eq:self1}) includes coupling with all bound trion states as well as unbound trion states. The latter are just exciton-electron scattering states and need to be treated carefully. The symmetric/anti-symmetric four-body wavefunction, with center of mass momentum $\vec{Q}$, of a state consisting of an exciton scattered from initial momentum $\vec{Q}$ to $\vec{Q}-\vec{q}_{o}$, and an electron scattered from initial momentum $\vec{p}_{o}$ inside the Fermi sea to momentum $\vec{p}_{o} + \vec{q}_{o}$ outside the Fermi sea, can be approximated as,
\begin{eqnarray}
  && \phi^{tr}_{m,\vec{Q}}(\vec{k}_{1},s_{1};\vec{k}_{2},s_{2};\vec{p},s_{2})  \approx \nonumber \\
  && \left[ \sqrt{\frac{A}{2}} \phi^{ex}_{n,\vec{Q}-\vec{q}_{o}}(\vec{k}_{1} + (\xi - \lambda_{e})\vec{Q} + \xi \vec{p}_{o} + \lambda_{e}\vec{q}_{o}) \right. \nonumber \\
    && \left. \times \delta_{\vec{k}_{2},(\xi + \eta)\vec{p}_{o} - \xi \vec{Q} + \vec{q}_{o}} \pm  \left\{ \vec{k}_{1} \leftrightarrow \vec{k}_{2}  \right\} \right] \nonumber \\
  & & \times \sqrt{A f_{c,s_{2}}(\vec{p}_{o})} \delta_{\vec{p}_{o},\vec{p}} \nonumber \\
  \label{eq:exelscat}
\end{eqnarray}
The summation over the index $m$ in (\ref{eq:self1}) would now involve a summation over all momenta $\vec{p}_{o}$ inside the Fermi sea and all transferred momenta $\vec{q}_{o}$, as well as all symmetric and anti-symmetric states consistent with the values of $s$ and $s'$. The energy of the above state, for small values of $\vec{Q}$ and $\vec{q}_{o}$, is,
\begin{eqnarray}
  & & E^{tr}_{m}(\vec{Q},s_{1},s_{2}) \rightarrow E^{ex}_{n}(\vec{Q}=0,s_{1},s_{2}) + \frac{\hbar^{2} q^{2}_{o}}{2 m_{T}} \nonumber \\
  & &  - \hbar^{2} \left(\frac{\vec{Q}}{m_{ex}} - \frac{\vec{p}_{o}}{m_{e}} \right) \cdot \vec{q}_{o} \nonumber \\ 
\end{eqnarray}
Here, $E^{ex}_{n}(\vec{Q}=0,s_{1},s_{2}) = 0.5(E^{ex}_{n}(\vec{Q}=0,s_{1}) + E^{ex}_{n}(\vec{Q}=0,s_{2}))$. $m_{T} = m_{ex}m_{e}/(m_{ex}+m_{e})$, is the reduced mass of the exciton and the electron. Use of the state in (\ref{eq:exelscat}) to evaluate exciton-electron scattering contributions to the self-energy in (\ref{eq:self1}) is equivalent to the use of the Born approximation in exciton-electron scattering ~\cite{Cohen03}. The contribution of exciton-electron scattering to the exciton self-energy is found to be,
\begin{eqnarray}
  && \Sigma^{ex}_{n,s}(\vec{Q},\omega)|_{ex-e} = \sum_{u,s'} \int \frac{d^{2}\vec{q}_{o}}{(2\pi)^{2}}  \int \frac{d^{2}\vec{p}_{o}}{(2\pi)^{2}} f_{c,s'}(\vec{p}_{o}) \nonumber \\
  & & \times \frac{(1 + \delta_{s,s'}) \left| h_{n,u}(\vec{Q},\vec{p}_{o},\vec{q}_{o},s,s')\right|^{2}}{\hbar \omega - \left[ E^{ex}_{n}(\vec{Q}=0,s,s') + \frac{\hbar^{2} q^{2}_{o}}{2 m_{T}} - \hbar^{2} \left(\frac{\vec{Q}}{m_{ex}} - \frac{\vec{p}_{o}}{m_{e}}  \right) + i\gamma_{tr} \right] } \nonumber \\
  \label{eq:exee}
  \end{eqnarray}
The summation over the variable $u = \pm 1$ involves a summation over all symmetric and antisymmetric unbound trion states (or exciton-electron scattering states) consistent with the values of $s$ and $s'$. The expression for $h_{n,u}(\vec{Q},\vec{p}_{o},\vec{q}_{o},s,s')$ is given in the Appendix.

\subsection{Optical Conductivity Sum Rule}
This optical conductivity sum rule for 2D TMDs can be derived from the restricted Thomas-Reiche-Kuhn optical conductivity sum rule~\cite{TRK98} and, assuming a full valence band, can be expressed as~\cite{TRK98},
\begin{equation}
  \int_{0}^{\infty} \omega {\rm Re}\{\sigma_{\vec{Q}}(\omega)\} \, \frac{d\omega}{2\pi} = \frac{e^{2}v^{2}}{2\hbar} \sum_{s} \int \frac{\displaystyle d^{2}\vec{k}}{\displaystyle (2\pi)^{2}} \left( 1 -  f_{c,s}(\vec{k}) \right) \label{eq:sum}
\end{equation}
Here, $v$ is the interband velocity matrix element between the valence and conduction band Bloch states (see Sec.~\ref{subsec:hamiltonian}). Band filling is incorporated into the above sum rule. The completeness of the exciton eigenfunctions $\phi^{ex}_{n,\vec{Q}}(\vec{k})$ can be used to see that the derived optically conductivity in (\ref{eq:cond1}) satisfies the above sum rule exactly. 

\section{Variational Eigenstates and Connection with Exciton-Polaron States}
Variational states for exciton-polarons have been constructed in previous works ~\cite{Imam16,Chang19}. These variational states resemble the Fermi polaron states of an impurity atom in a cold Fermi gas ~\cite{Demler12,Demler18,Chevy06}. Here we show that variational states can be constructed using the eigenstates of the two-body and four-body Schr{\"o}dinger equations in (\ref{eq:exciton1}) and (\ref{eq:4body1}), respectively, and which give results for the eigenenergies in exact agreement with the exciton self-energy given in (\ref{eq:self1}). The eigenenergies can be obtained from the poles of the exciton Green's function and these energies are the roots of the equation,
\begin{equation}
  \hbar\omega - E^{ex}_{n}(\vec{Q},s) + i\gamma_{ex} - \Sigma^{ex}_{n,s}(\vec{Q},\omega) = 0 \label{eq:roots}
\end{equation}
where the exciton self-energy is as given in (\ref{eq:self1}). Since the two-body and four-body Schr{\"o}dinger equations are coupled via the Coulomb matrix elements, one can construct approximate eigenstates of the Hamiltonian (within the purview of single CB electron-hole pair excitations) by a simple superposition as follows,
\begin{eqnarray}
  & &  |\psi_{n,s}(\vec{Q})\rangle = \frac{\alpha_{n}}{\sqrt{A}} \sum_{k} \frac{\phi^{ex*}_{n,\vec{Q}}(\vec{k})}{N_{ex}}  \nonumber \\
  & & \times c^{\dagger}_{s}(\vec{k}+\lambda_{e}\vec{Q}) b_{s}(\vec{k}-\lambda_{h}\vec{Q}) |GS \rangle \nonumber \\
  && + \sum_{m,s'} \frac{\beta_{m}}{\sqrt{A^{3}}} \sum^{\vec{\underline{k}}_{1},\vec{\underline{k}}_{2} \ne \vec{p}}_{\vec{k}_{1},\vec{k}_{2},\vec{p}} \frac{\phi^{tr*}_{m,\vec{Q}}(\vec{k}_{1},s;\vec{k}_{2},s';\vec{p},s')}{N_{tr}} \nonumber \\
  & & \times \, c^{\dagger}_{s}(\vec{\underline{k}}_{1}) c^{\dagger}_{s'}(\vec{\underline{k}}_{2}) b_{s}(\vec{\underline{k}}_{1} + \vec{\underline{k}}_{2}-(\vec{Q}+\vec{p})) c_{s'}(\vec{p}) |GS \rangle \nonumber \\
  \label{eq:var}
\end{eqnarray}
where $|GS \rangle$ is the ground state of the electron doped material. The above state resembles a Fermi-polaron variational state~\cite{Imam16,Demler12,Chevy06,Demler18}. The normalization terms are,
\begin{eqnarray}
  & & N_{ex} = \sqrt{1-f_{c,s}(\vec{k}+\lambda_{e} \vec{Q})} \nonumber \\
  & & N_{tr} = \sqrt{(1 + \delta_{s,s'}) f_{c,s'}(\vec{p}) \left[ 1-f_{c,s}(\vec{\underline{k}}_{1}) \right]\left[ 1 - f_{c,s'}(\vec{\underline{k}}_{2}) \right]} \nonumber \\
  \end{eqnarray}
The underlined vectors, $\vec{\underline{k}}_{1}$ and $\vec{\underline{k}}_{2}$, are defined as earlier in Section~\ref{sec:eom}. The states in the superposition are properly normalized and are orthogonal. The restrictions $\vec{\underline{k}}_{1},\vec{\underline{k}}_{2} \ne \vec{p}$ in the summations in the second term follow from the fact that the states of the four-body Schr{\"o}dinger equation correspond to fully connected diagrams and should have no direct optical matrix element with the ground state $|GS \rangle$. This restriction also keeps the superposed trion states in the variational state orthogonal to the exciton states. When the variational state given above is used to minimize energy with respect to the Hamiltonian given earlier in (\ref{eq:H}), the eigenfunctions $\phi^{tr}_{m,\vec{Q}}(\vec{k}_{1},s;\vec{k}_{2},s';\vec{p},s')$ and $\phi^{ex}_{n,\vec{Q}}(\vec{k})$, as expected, are found to obey the coupled two-body and four-body Schr{\"o}dinger equations, the trion states are found to be coupled to the exciton states via the Coulomb matrix elements $M_{m,n}(\vec{Q},s,s')$ given earlier (see (\ref{eq:M})), and the energy eigenvalues $\hbar \omega$ are found to obey (\ref{eq:roots}) provided $\gamma_{ex}$ and $\gamma_{tr}$ are set to zero. Therefore, the same physics is captured by our coupled two-body and four-body Schr{\"o}dinger equations and the exciton-polaron formalism. Furthermore, the optical conductivity calculated using the above variational state also matches the one found earlier in (\ref{eq:cond1}). The formalism presented here shows that the variational polaron state can be written in terms of the eigenstates of the two-body and the four-body Schr{\"o}dinger equations (i.e. in terms of the exciton and trion eigenstates) and that all bound and unbound trion states must be included in the variational polaron state. Signatures of the resulting quantum coherence between the exciton and trion states have been observed experimentally~\cite{Hao16}.

\section{Numerical Simulation Results and Discussion}
For simulations, we consider a monolayer of 2D MoSe$_{2}$ on a SiO$_{2}$ substrate. In monolayer MoSe$_{2}$, spin-splitting of the conduction bands is large ($\sim$35 meV~\cite{Kosmider13}) and the lowest conduction band in each of the $K$ and $K'$ valleys is optical coupled to the topmost valence band~\cite{Xiao13}. We use effective mass values of $0.7 m_{o}$ for both $m_{e}$ and $m_{h}$ which agree with the recently measured value of $0.35 m_{o}$ for the exciton reduced mass~\cite{Goryca19}. We use a wavevector-dependent dielectric constant $\epsilon(\vec{q})$ for the Coulomb potentials appropriate for 2D materials, as described in our earlier work~\cite{Changjian14}, to screen the Coulomb potentials. We should emphasize here that besides $\gamma_{ex}$ and $\gamma_{tr}$, and unlike in previous works~\cite{Imam16,Macdonald17}, there are no other free parameters in our theoretical model and no artificial upper cut-offs of momenta integrals to avoid divergences.

\begin{figure}
  \begin{center}
    \includegraphics[width=0.8\columnwidth]{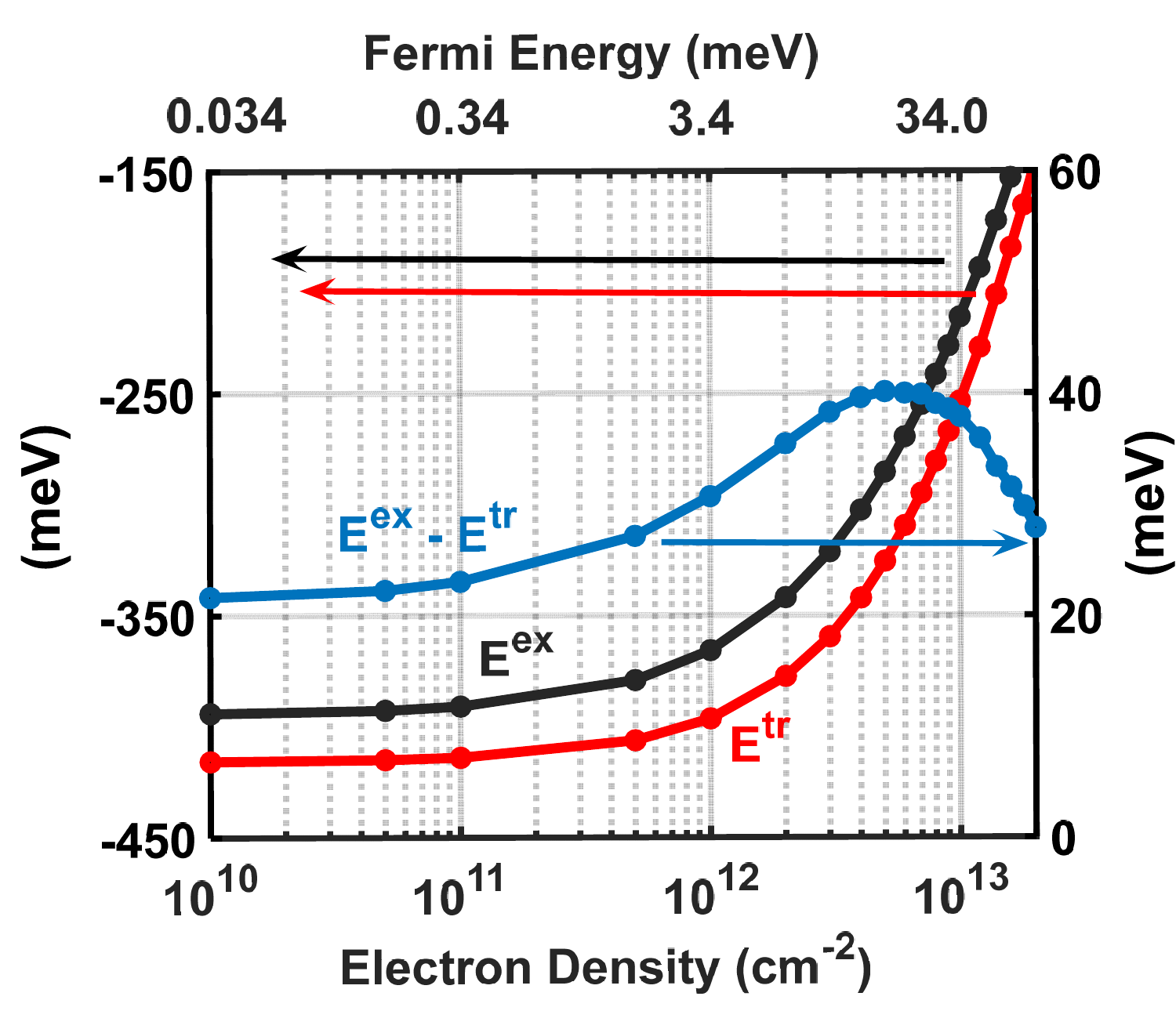}
    \caption{Calculated energies, $E^{tr}_{m=0}(\vec{Q}=0,s_{1},s_{2})$ and $E^{ex}_{n=0}(\vec{Q}=0,s_{1})$, of the lowest energy bound trion and exciton states, respectively, referenced to the material bandgap, are plotted as a function of the electron density (and Fermi energy) for monolayer 2D MoSe$_{2}$ on SiO$_{2}$. Trion binding energy $E^{tr}_{b}$ is also plotted. T = 5K.}
    \label{fig:fig3}
  \end{center}
\end{figure}
\begin{figure}
  \begin{center}
   \includegraphics[width=0.70\columnwidth]{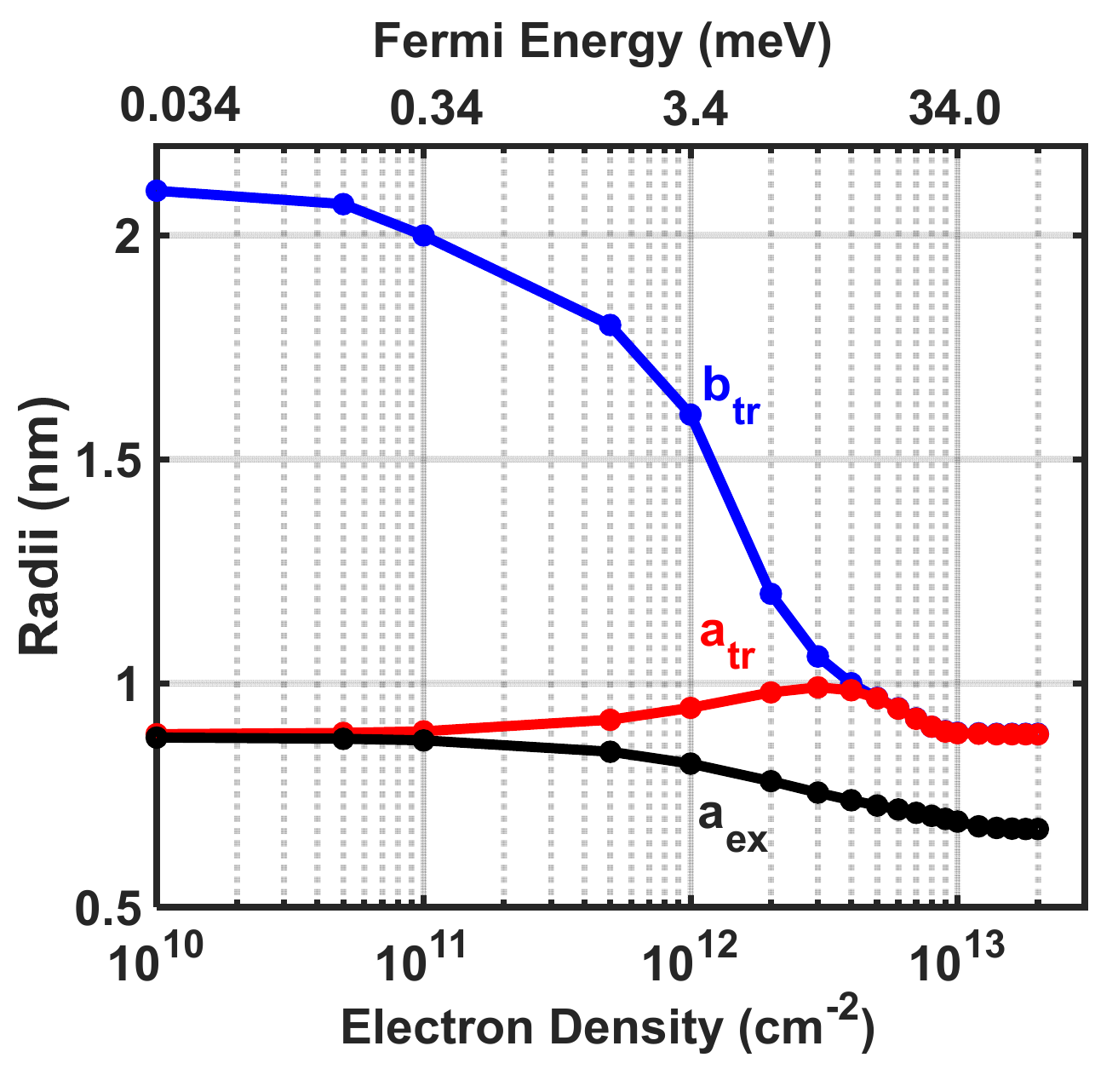}
   \caption{ Calculated trion radii, $a_{tr}$ and $b_{tr}$, for the variational trion wavefunction, and the exciton radius $a_{ex}$ for the variational exciton wavefunction are plotted as a function of the electron density (and Fermi energy) for monolayer 2D MoSe$_{2}$ on SiO$_{2}$. T = 5K.}   
    \label{fig:fig4}
  \end{center}
\end{figure}

\subsection{Trion and Exciton Radii and Energies}
As discussed in previous works~\cite{Changjian14,Berk13,Macdonald17,Chang19,Imam16}, exchange correlations favor only singlet trion bound states in MoSe$_{2}$, in which the exciton belongs to one valley and the bound electron-hole pair belongs to the other valley. We write a product variational wavefunction for the lowest energy four-body bound singlet trion eigenstate with $\vec{Q=0}$, as follows,
\begin{equation}
  \phi^{tr}_{m=0,\vec{Q}=0}(\vec{k}_{1},s_{1};\vec{k}_{2},s_{2};\vec{p},s_{2}) = \chi(\vec{k}_{1},s_{1};\vec{k}_{2},s_{2}) \varphi(\vec{p},s_{2})
  \end{equation}
We assume the following symmetric variational wavefunction for $\chi(\vec{r}_{1},s_{1};\vec{r}_{2},s_{2})$ (assuming $s_{1} \ne s_{2}$)~\cite{Berk13,Changjian14,Urba17},  
\begin{eqnarray}
  & & \chi(\vec{r}_{1},s_{1};\vec{r}_{2},s_{2}) \propto  \left[ e^{-|\vec{r}_{1}|/a_{tr} - |\vec{r}_{2}|/b_{tr}}  + \{ \vec{r}_{1} \leftrightarrow \vec{r}_{2} \} \right] \nonumber \\
\end{eqnarray}
The trion radii, $a_{tr}$ and $b_{tr}$, are variational parameters. The form of the CB hole wavefunction $\varphi(\vec{p},s_{2})$ is chosen so as to minimize Coulomb energy and maximize coupling between the trion and exciton states, as discussed in Sec.~\ref{sec:hole}. Since the trion radii are expected to be much smaller than the size of the CB hole, which cannot be smaller than $\sim1/k_{F}$, the simplest and the easiest way to get the smallest CB hole is to assume that $\varphi(\vec{p},s_{2}) = \sqrt{f_{c,s_{2}}(\vec{p})/n_{s_{2}}}$, where proper wavefunction normalization requires $n_{s_{2}} = A^{-1}\sum_{\vec{p}} f_{c,s_{2}}(\vec{p})$. For the lowest energy $\vec{Q}=0$ bound exciton state we use the variational wavefunction~\cite{Berk13,Changjian14,Urba17},
\begin{equation}
\phi^{ex}_{n=0,\vec{Q}=0}(\vec{r}) \propto  e^{-|\vec{r}|/a_{ex}} 
\end{equation}
Using the radii, $a_{tr}$, $b_{tr}$, and $a_{ex}$, as variational parameters, we find the eigenenergies, $E^{tr}_{m=0}(\vec{Q}=0,s_{1},s_{2})$ and $E^{ex}_{n=0}(\vec{Q}=0,s_{1})$, as a function of the electron density. The results are shown in Fig.~\ref{fig:fig3} which plots these energies with respect to the material bandgap $E_{g}$. The corresponding trion and exciton radii are plotted in Fig.~\ref{fig:fig4}. The trion binding energy $E^{tr}_{b}$, defined as $E^{ex}_{n=0}(\vec{Q}=0,s_{1})-E^{tr}_{m=0}(\vec{Q}=0,s_{1},s_{2})$, is also plotted. The exciton binding energy decreases with the electron density due to phase space filling~\cite{Changjian14}. The trion binding energy first increases with the electron density and then it decreases. The initial increase is due to two reasons,
\begin{itemize}
\item Suppose an exciton with center of mass momentum $\vec{Q}=0$ grabs an electron with momentum $\vec{p}$ within the Fermi sea to form a four-body bound state with momentum $\vec{Q}=0$. The center of mass kinetic energy of the four-body state would be $\hbar^{2}p^{2}/(2m_{t}) - \hbar^{2}p^{2}/(2m_{e})$. The first term is the center of mass kinetic energy of the two CB electrons and one VB hole in the four-body bound state. The second term represents the kinetic energy of the CB hole in the four-body bound state. These energies are included in the terms on the LHS of (\ref{eq:4body1}). Averaging this energy with respect to the CB hole wavefunction $\varphi(\vec{p},s)$ contributes a factor $E_{F}/3$ to the trion binding energy.
\item At small electron densities, phase space filling restricts electron-electron Coulomb repulsion more than electron-hole Coulomb attraction.
\end{itemize}

As the electron density increases further, the reduced phase space diminishes electron-hole Coulomb attraction as well and the binding energy of the trion decreases rapidly. It remains an open question if the trion binding energy eventually goes to zero or not at high enough electron densities. The approximations made in this work do not permit us to generate reliable results for electron densities higher than $2\times10^{13}$ cm$^{-2}$. Interestingly, the Fermi energy remains smaller than the trion binding energy for electron densities smaller than $\sim 10^{13}$ 1/cm$^{3}$. Note that the binding energies of the exact energy eigenfunctions are expected to be larger than our variational solutions. In this work, no bound trions states with an anti-symmetric wavefunction were found even for vanishingly small electron densities. In addition, no bound trion states were found in which the CB hole was not bound to the two CB electrons and the VB hole. 

Fig.~\ref{fig:fig4} shows that the exciton and the larger trion radii decrease with the electron density because phase space blocking inside the Fermi surface causes the wavefunctions to spread out more in the momentum space~\cite{Changjian14}. For electron densities higher than $\sim 5\times 10^{12}$ cm$^{-2}$, the two trion radii are almost identical and approximately equal to 0.9 nm.

\begin{figure}
  \begin{center}
   \includegraphics[width=0.8\columnwidth]{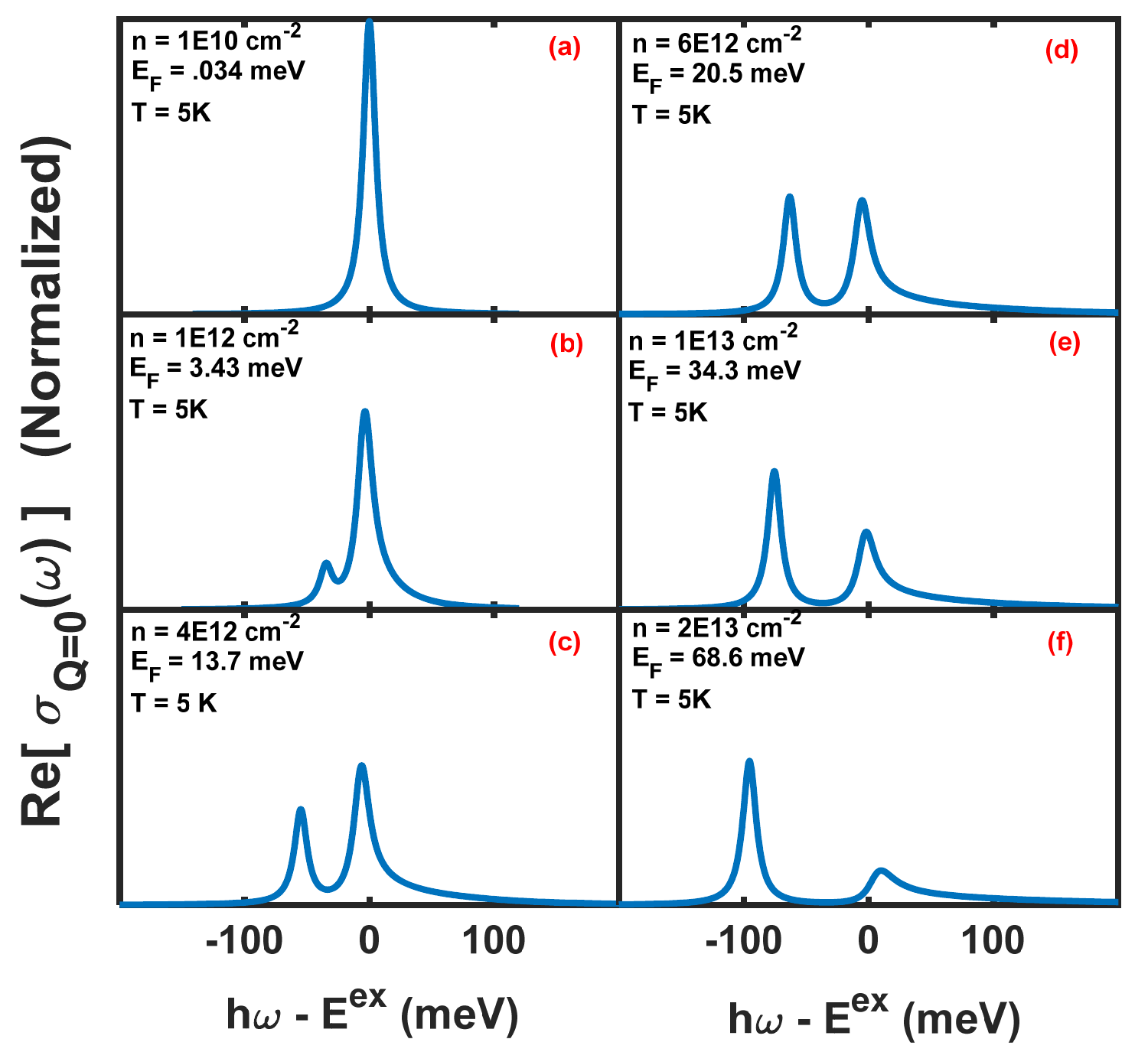}
   \caption{Calculated real part of the optical conductivity, $\sigma_{\vec{Q}=0}(\omega)$, is plotted for different electron densities for monolayer 2D MoSe$_{2}$ on SiO$_{2}$. The spectra are all normalized to peak optical conductivity value at zero electron density. T = 5K. The frequency axis is offset by the exciton eigenenergy $E^{ex}_{0}(\vec{Q},s)$ of the two-body Schr{\"o}dinger equation.}   
    \label{fig:fig5}
  \end{center}
\end{figure}

\begin{figure}
  \begin{center}
   \includegraphics[width=0.7\columnwidth]{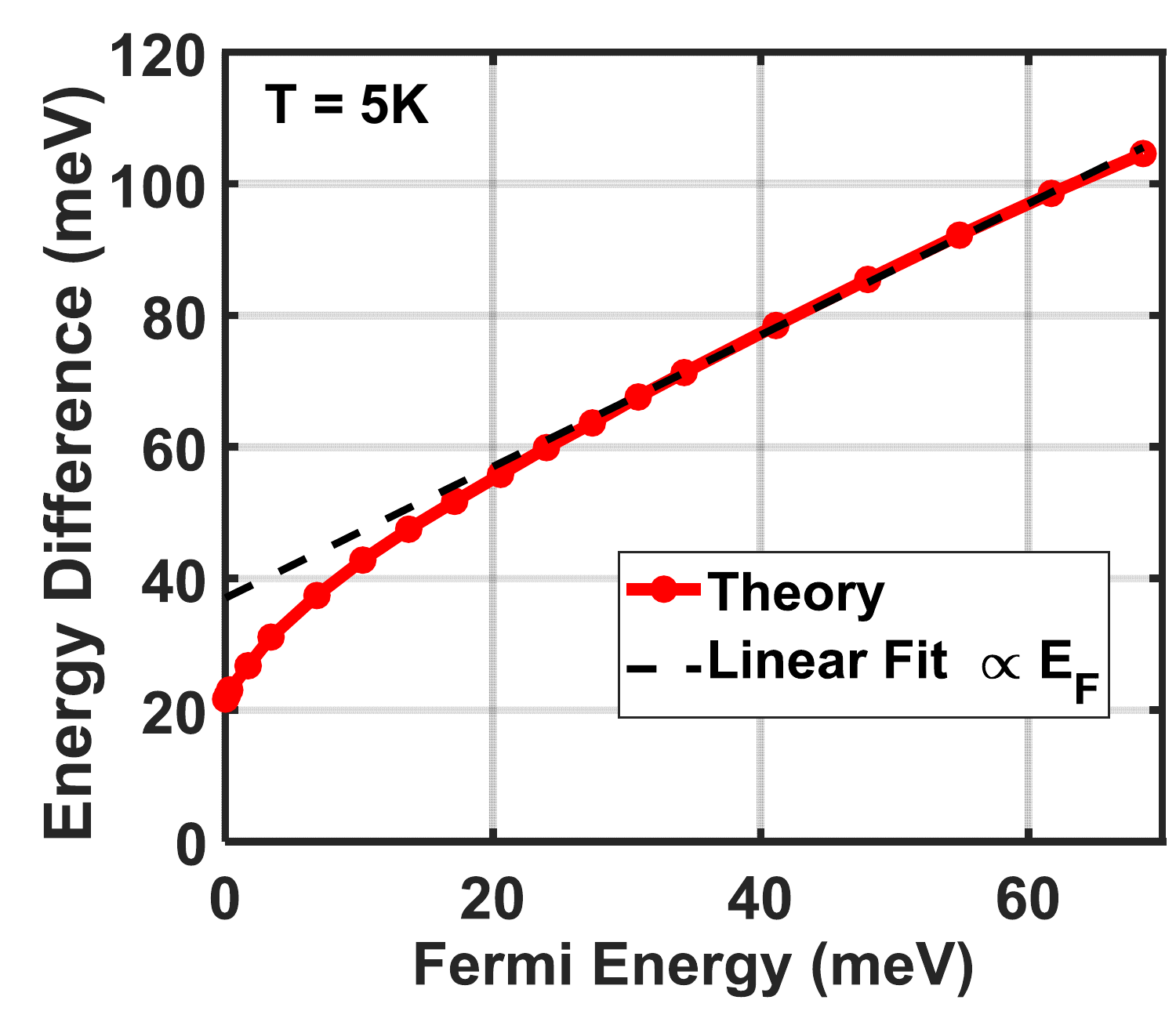}
   \caption{Calculated energy difference between the two dominant peaks in the optical absorption spectra in Fig.~\ref{fig:fig5} is plotted as a function of the Fermi energy $E_{F}$ for monolayer 2D MoSe$_{2}$ on SiO$_{2}$. The dashed line has unit slope and shows that the calculated energy difference varies as $E_{F}$ at high electron densities.}   
    \label{fig:fig6}
  \end{center}
\end{figure}

\begin{figure}
  \begin{center}
   \includegraphics[width=0.8\columnwidth]{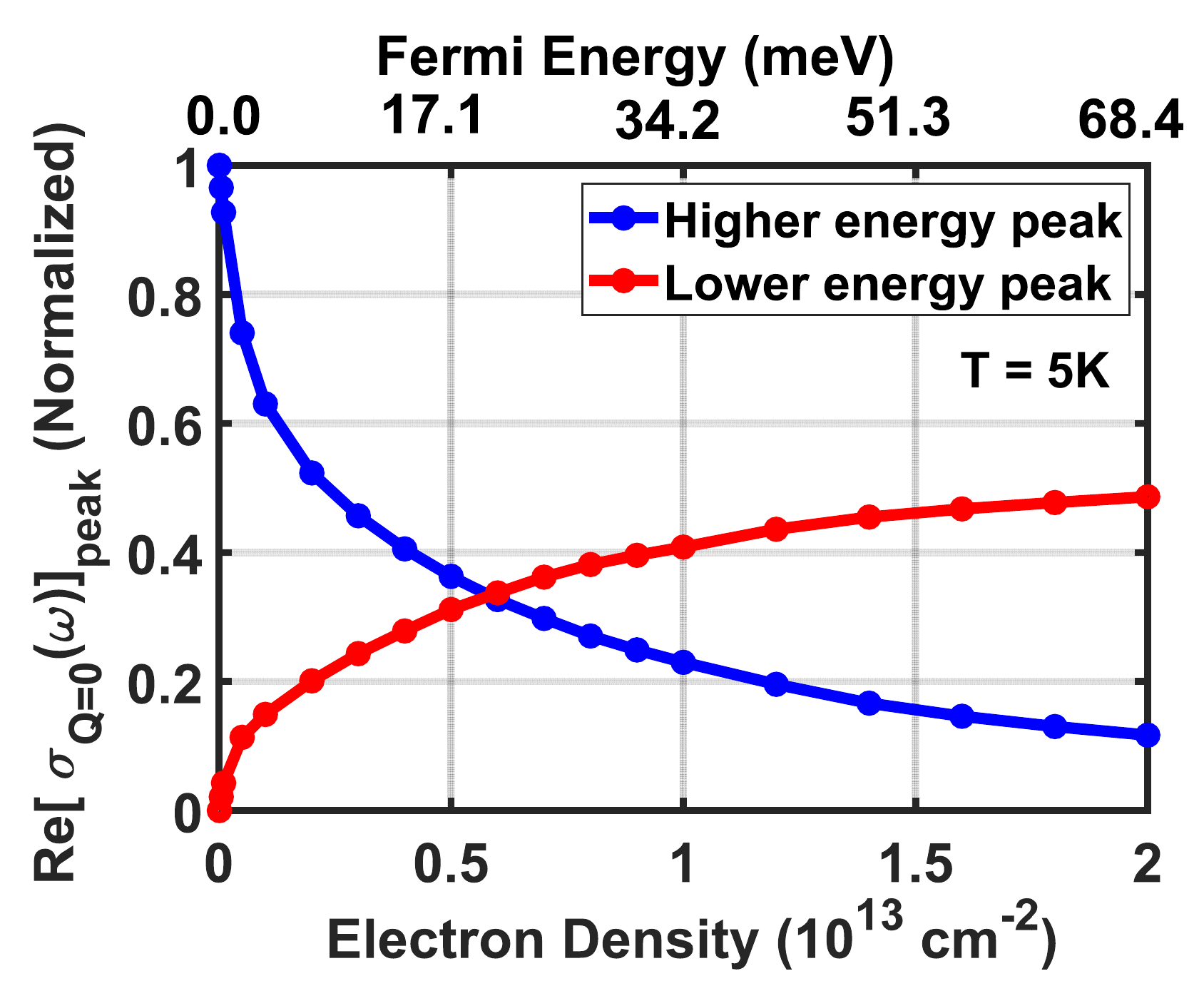}
   \caption{The maximum optical conductivity (real part) values of the two absorption peaks in Fig.~\ref{fig:fig5}, normalized to the maximum optical conductivity value at zero electron density, are plotted as a function of the electron density for monolayer 2D MoSe$_{2}$ on SiO$_{2}$.}   
    \label{fig:fig7}
  \end{center}
\end{figure}

\subsection{Optical Conductivity Spectra}
Fig.~\ref{fig:fig5} shows the calculated real part of the optical conductivity $\sigma_{\vec{Q}=0}(\omega)$ plotted for different electron densities for monolayer 2D MoSe$_{2}$ on SiO$_{2}$. The expression in (\ref{eq:cond1}) has been used to generate the plots in Fig.~\ref{fig:fig5}. In our optical conductivity calculations, we have included only the lowest energy bound exciton and bound trion states, as well as the unbound trion states that describe exciton-electron scattering for the lowest energy bound exciton state in the Born approximation, as discussed in Sec.~\ref{sec:ees}. The values of $\gamma_{ex}$ and $\gamma_{tr}$ were both chosen to be 4 meV.    

\begin{figure}
  \begin{center}
   \includegraphics[width=0.75\columnwidth]{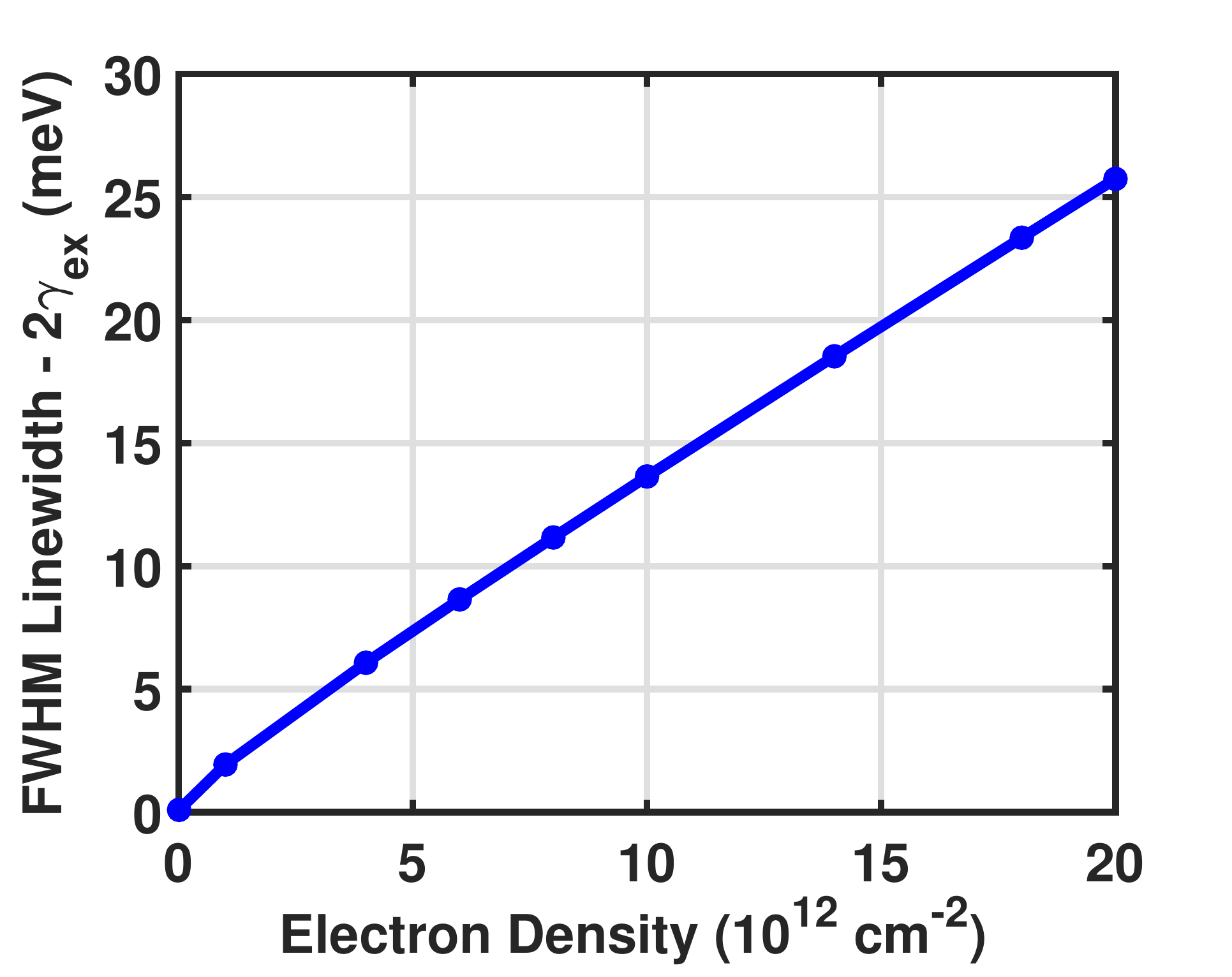}
   \caption{The (FWHM) linewidth (with $2\gamma_{ex}$ subtracted) of the higher energy absorption peak in Fig.~\ref{fig:fig5} is plotted as a function of the electron density for monolayer 2D MoSe$_{2}$ on SiO$_{2}$. The plotted linewidth with $2\gamma_{ex}$ subtracted displays the linewidth resulting from exciton-electron interactions. The linewidth of the lower energy absorption peak in Fig.~\ref{fig:fig5} does not change with the electron density.}   
    \label{fig:fig9}
  \end{center}
\end{figure}

The spectra shows two distinct peaks. At small electron densities, the higher energy peak dominates and carries all the spectral weight. As the electron density increases, the spectral weight shifts from the higher energy peak to the lower energy peak. The higher and lower energy peaks have also been called repulsive and attractive exciton-polaron peaks, respectively, by Efimkin et al.~\cite{Macdonald17}. Very often in the literature they are just referred to as the exciton peaks and the trion peak, respectively~\cite{Fai13,Chernikov15}. These two peaks arise from the Coulomb coupling of the excitons and the trions, as discussed earlier in this paper. For very small electron densities, their energies coincide with those of excitons and trions as obtained from the two-body and the four-body Schr{\"o}dinger equations. The coupling between the excitons and the trions, described by the matrix elements $M_{m,n}(\vec{Q},s,s')$ in (\ref{eq:M}), increases with the electron density and, therefore, the energy difference between the two peaks in the absorption spectra also increases with the electron density. Fig.~\ref{fig:fig6} plots this energy difference as a function of the Fermi energy $E_{F}$ for monolayer 2D MoSe$_{2}$ on SiO$_{2}$. The dashed line has unit slope and shows that the calculated energy difference varies approximately as $E_{F}$ at high electron densities. Fig.~\ref{fig:fig5} shows that as the electron density increases, the lower energy peak shifts down to lower energies much more than the upward motion of the higher energy peak. This happens because the continuum of exciton-electron scattering states lies just above the higher energy peak and prevent the higher energy peak from moving upwards too much. Fig.~\ref{fig:fig7} plots the peak optical conductivity (real part) of the two absorption peaks, normalized to the peak optical conductivity at zero electron density, as a function of the electron density. The lower energy peak begins to dominate when the electron density exceeds $6 \times 10^{12}$ cm$^{-2}$. As the electron density increases, the higher energy peak also loses spectral weight to the broad continuum of single electron-hole pair excitations from exciton-electron scattering. This results not only in the broadening of the higher energy peak but also in the appearance of a broad pedestal around the base of the peak that is more prominent on its higher energy side. The lower energy peak, on the other hand, does not broaden as the electron density increases. Fig.\ref{fig:fig9} shows the (FWHM) linewidth (with $2\gamma_{ex}$ subtracted) of the higher energy absorption peak plotted as a function of the electron density for monolayer 2D MoSe$_{2}$ on SiO$_{2}$. The plotted linewidth with $2\gamma_{ex}$ subtracted displays the linewidth resulting from just exciton-electron interactions. The contribution to the linewidth from exciton-electron interactions increases almost linearly with the electron density (and the Fermi energy).

The integrated area under the plotted conductivity spectra in Fig.~\ref{fig:fig5} is almost conserved (but not exactly conserved) as the electron density increases, in agreement with the sum rule in (\ref{eq:sum}). These observations are all in good agreement with the experimental results reported for two dimensional materials~\cite{Fai13,Chernikov15}, and with previous theoretical works~\cite{Macdonald17,Chang19,Suris01b}.   

\section{Conclusion}
In this paper, we presented a theoretical model that explains the behavior of trions and excitons in doped 2D materials. Coulomb scattering couples the exciton and trion states in doped materials. This coupling is well described by two coupled Schr{\"o}dinger equations for excitons and trions that we derived using the many body density matrix technique. The calculated optical conductivity was shown to explain the prominent features of the experimentally measured optical absorption spectra and also satisy the optical conductivity sum rule exactly. The eigensolution of the coupled Schr{\"o}dinger equations, constructed using a superposition of exciton and trion states, had the same form as a Fermi polaron state and revealed the connection between our approach and polaron physics. The work presented here will help to clarify the physics associated with excitons and trions in doped 2D materials.

There are still several questions that remain open in this area. For example, it is not clear if the trion eigenequation has bound states at very high electron densities (much larger than $\sim 10^{13}$ cm$^{-2}$). If not, then how does the conductivity spectra evolve at very high electron densities? At high enough electron densities, multiple electron-hole pair excitations, ignored in this work, are also expected to become important. Their inclusion is expected to broaden the lower energy absorption peak as well and steal spectral weight from it. Indeed, multiple pair excitations have already been shown to play an important role in Fermi polaron physics in atomic systems~\cite{Parish13}. Finally, the role of Fermi edge affects, which involves multiple electron-hole pair excitations, in this context is not clear but they are also expected to become increasingly important at high electron densities. Exploring answers to these questions will be the subject of future work.    

\section{Acknowledgments}
The authors would like to acknowledge helpful discussions with Nick Vamivakas, Minwoo Jung, Francesco Monticone, Jacob Khurgin, and Gennady Shvets, and support from CCMR under NSF grant number DMR-1120296, ONR under grant number N00014-12-1-0072, and NSF EFRI-NewLaw under grant number 1741694. 

\appendix{}

\section{Appendix: Hermitian Two-Body Exciton Schr{\"o}dinger Equation}  \label{app1}
Equation (\ref{eq:exciton1}), without the first term and the last two terms on the RHS, is the standard eigenvalue equation for excitons~\cite{haugbook,Kira12}. However, the equation is not Hermitian. It can be converted into a Hermitian equation. We define $\overline{P}_{\vec{Q}}(\vec{k},s)$ as $P_{\vec{Q}}(\vec{k},s)/\sqrt{1 - f_{c,s}(\vec{k} + \vec{Q})}$. We also include $\left[ 1 - f_{c,s}(\vec{k}+\vec{Q}) \right]$ in the last term on the RHS in (\ref{eq:exciton1}) containg the Coulomb potential $V(\vec{q})$. This added factor does not show up at this level in the density matrix technique but its inclusion ensures the Hermiticity of the set of coupled two-body and four-body Schr{\"o}dinger equations. Physically, it restricts the phase space for electron scattering just like the first and the second terms on the RHS. With these definitions and changes, we obtain,  
\begin{eqnarray}
& &  \left[ {\bf E}_{c,s}(\vec{k} + \vec{Q}) - {\bf E}_{v,s}(\vec{k}) +i\gamma_{ex} - \hbar \omega \right]  \overline{P}_{\vec{Q}}(\vec{k},s)= \nonumber \\
& & - \frac{1}{\sqrt{A}} g^{*}_{s} \langle a^{\dagger}(\vec{Q}) \rangle \sqrt{1 - f_{c,s}(\vec{k}+\vec{Q})} \nonumber \\
  & & + \frac{\sqrt{1 - f_{c,s}(\vec{k}+\vec{Q})}}{A}\sum_{\vec{q}} U(\vec{q}) \nonumber \\
  & & \times \overline{P}_{\vec{Q}}(\vec{k} + \vec{q},s)  \sqrt{1 - f_{c,s}(\vec{k} + \vec{Q} + \vec{q})} \nonumber \\
& & -\frac{1}{A} \sum_{\vec{q},\vec{p},s'} U(\vec{q}) \nonumber \\
& & \times \frac{T^{c}_{\vec{Q}}(\vec{k} + (\xi + \eta)\vec{Q} - \xi \vec{p},s;(\xi + \eta)\vec{p} - \xi \vec{Q} - \vec{q},s';\vec{p},s')}{\sqrt{1 - f_{c,s}(\vec{k}+\vec{Q})}} \nonumber \\
& & +\frac{1}{A} \sum_{\vec{q},\vec{p},s'} V(\vec{q}) \sqrt{1 - f_{c,s}(\vec{k}+\vec{Q})} \nonumber \\
& & \times T^{c}_{\vec{Q}}(\vec{k} + (\xi + \eta)\vec{Q} - \xi \vec{p} +\vec{q},s;(\xi + \eta)\vec{p} - \xi \vec{Q} - \vec{q},s';\vec{p},s') \nonumber \\
  \label{eq:excitonapp1}
\end{eqnarray}
The homogeneous part of the above equation is now a Hermitian eigenvalue equation. It has a complete set of orthonormal eigenfunctions $\phi^{ex}_{n,\vec{Q}}(\vec{k}+\lambda_{h}\vec{Q})$. In the limit of very low electron density, when phase space filling effects can be ignored, and assuming $Q<<k_{F}$, the eigenenergies of the bound exciton states can be expressed as,
\begin{equation}
  E^{ex}_{n}(\vec{Q},s) = E^{ex}_{n}(\vec{Q}=0,s) + \frac{\hbar^{2}Q^{2}}{2 m_{ex}}
\end{equation}

\section{Appendix: Hermitian Four-Body Trion Schr{\"o}dinger Equation}  \label{app2}
Equation (\ref{eq:4body1}), without the last two terms on the RHS, is a four-body eigenvalue equation for trions. The equation is not Hermitian. It can be converted into a Hermitian eigenvalue equation with a few approximations. The term $\left[ 1-f_{c,s_{1}}(\vec{\underline{k}}_{1}) - f_{c,s_{2}}(\vec{\underline{k}}_{2})  \right]$ on the RHS can be replaced by $\left[ 1-f_{c,s_{1}}(\vec{\underline{k}}_{1}) \right]\left[ 1 - f_{c,s_{2}}(\vec{\underline{k}}_{2}) \right]$. The difference between the two, $-f_{c,s_{1}}(\vec{\underline{k}}_{1}) f_{c,s_{2}}(\vec{\underline{k}}_{2})$, stems from the fact that the four-body correlation function $T^{c}_{\vec{Q}}$ can be non-zero if the correlations are between electrons outside the Fermi sea or if they are between holes inside the Fermi sea. In this work, correlation between holes in the Fermi sea may be ignored since the trion radii are smaller than the inverse Fermi momentum for electron densities smaller than $2\times 10^{13}$ cm$^{-2}$. Similarly, one can replace the terms $\left[ f_{c,s_{2}}(\vec{p}) - f_{c,s_{1/2}}(\vec{\underline{k}}_{1/2}) \right]$ on the RHS by  $f_{c,s_{2}}(\vec{p}) \left[ 1 - f_{c,s_{1/2}}(\vec{\underline{k}}_{1/2}) \right]$. We then define $\overline{T}^{c}_{\vec{Q}}(\vec{k}_{1},s_{1};\vec{k}_{2},s_{2};\vec{p},s_{2})$ as,
\begin{eqnarray}
  & &  \overline{T}^{c}_{\vec{Q}}(\vec{k}_{1},s_{1};\vec{k}_{2},s_{2};\vec{p},s_{2}) = \nonumber \\
  & & \frac{T^{c}_{\vec{Q}}(\vec{k}_{1},s_{1};\vec{k}_{2},s_{2};\vec{p},s_{2})}{  \sqrt{f_{c,s_{2}}(\vec{p}) \left[ 1-f_{c,s_{1}}(\vec{\underline{k}}_{1}) \right]\left[ 1 - f_{c,s_{2}}(\vec{\underline{k}}_{2})  \right]}} 
\end{eqnarray}
With the above approximations and definitions, we obtain,
\begin{eqnarray}
  & & \left[ {\bf E}_{c,s_{1}}(\vec{\underline{k}}_{1}) + {\bf E}_{c,s_{2}}(\vec{\underline{k}}_{2}) - {\bf E}_{v,s_{1}}(\vec{\underline{k}}_{1}+\vec{\underline{k}}_{2} - (\vec{Q}+\vec{p})) \right. \nonumber \\
  & &  \left. - {\bf E}_{c,s_{2}}(\vec{p}) + i\gamma_{tr} - \hbar \omega \right] \overline{T}^{c}_{\vec{Q}}(\vec{k}_{1},s_{1};\vec{k}_{2},s_{2};\vec{p},s_{2}) =  \nonumber \\ 
  & & - \frac{\sqrt{\left[ 1-f_{c,s_{1}}(\vec{\underline{k}}_{1}) \right]\left[ 1 - f_{c,s_{2}}(\vec{\underline{k}}_{2}) \right] }}{A} \sum_{\vec{q}} V(\vec{q}) \times \nonumber \\  
  & & \overline{T}^{c}_{\vec{Q}}(\vec{k}_{1}+\vec{q},s_{1};\vec{k}_{2}-\vec{q},s_{2};\vec{p},s_{2}) \times \nonumber \\
  & & \sqrt{\left[ 1-f_{c,s_{1}}(\vec{\underline{k}}_{1} + \vec{q}) \right] \left[ 1 - f_{c,s_{2}}(\vec{\underline{k}}_{2} - \vec{q}) \right] } \nonumber \\
  & & + \frac{\sqrt{1-f_{c,s_{1}}(\vec{\underline{k}}_{1}) }}{A} \sum_{\vec{q}} U(\vec{q}) \overline{T}^{c}_{\vec{Q}}(\vec{k}_{1}+\vec{q},s_{1};\vec{k}_{2},s_{2};\vec{p},s_{2}) \times \nonumber \\
  & & \sqrt{1-f_{c,s_{1}}(\vec{\underline{k}}_{1} + \vec{q}) } \nonumber \\
  & & + \frac{\sqrt{1 - f_{c,s_{2}}(\vec{\underline{k}}_{2}) }}{A} \sum_{\vec{q}} U(\vec{q}) \overline{T}^{c}_{\vec{Q}}(\vec{k}_{1},s_{1};\vec{k}_{2}-\vec{q},s_{2};\vec{p},s_{2}) \times \nonumber \\
  & & \sqrt{1-f_{c,s_{2}}(\vec{\underline{k}}_{2} - \vec{q})} \nonumber \\
  & & + \frac{\sqrt{f_{c,s_{2}}(\vec{p}) \left[ 1- f_{c,s_{1}}(\vec{\underline{k}}_{1})  \right]}}{A} \times \nonumber \\
  & & \sum_{\vec{q}} V(\vec{q}) \overline{T}^{c}_{\vec{Q}}(\vec{k}_{1}+ (\xi + \eta)\vec{q},s_{1};\vec{k}_{2}-\xi \vec{q},s_{2};\vec{p}+\vec{q},s_{2}) \times \nonumber \\
  & & \sqrt{f_{c,s_{2}}(\vec{p}+\vec{q}) \left[ 1- f_{c,s_{1}}(\vec{\underline{k}}_{1}+\vec{q})  \right]} \nonumber \\
  & & + \frac{\sqrt{f_{c,s_{2}}(\vec{p}) \left[ 1- f_{c,s_{2}}(\vec{\underline{k}}_{2})  \right]}}{A} \times \nonumber \\
  & & \sum_{\vec{q}} V(\vec{q}) \overline{T}^{c}_{\vec{Q}}(\vec{k}_{1} + \xi \vec{q},s_{1};\vec{k}_{2} - (\xi + \eta)\vec{q},s_{2};\vec{p}-\vec{q},s_{2}) \times \nonumber \\
  & & \sqrt{f_{c,s_{2}}(\vec{p}-\vec{q}) \left[ 1- f_{c,s_{2}}(\vec{\underline{k}}_{2}-\vec{q})\right]} \nonumber \\
  & & - \frac{\sqrt{f_{c,s_{2}}(\vec{p})}}{A} \times \nonumber \\
  & & \sum_{\vec{q}} U(\vec{q}) \overline{T}^{c}_{\vec{Q}}(\vec{k}_{1}-\xi \vec{q},s_{1};\vec{k}_{2}- \xi \vec{q},s_{2};\vec{p}+\vec{q},s_{2}) \times \nonumber \\
  & & \sqrt{f_{c,s_{2}}(\vec{p}+\vec{q})} \nonumber \\
  & & + \frac{\sqrt{f_{c,s_{2}}(\vec{p}) \left[ 1-f_{c,s_{1}}(\vec{\underline{k}}_{1}) \right] \left[ 1 - f_{c,s_{2}}(\vec{\underline{k}}_{2}) \right]} }{A} \sum_{\vec{q}} V(\vec{q})  \times \nonumber \\
  & & \left[ P_{\vec{Q}}(\vec{\underline{k}}_{1} - \vec{Q} + \vec{q},s_{1}) \delta_{\vec{\underline{k}}_{2}-\vec{q},\vec{p}} \right. \nonumber \\
  & & \left. - P_{\vec{Q}}(\vec{\underline{k}}_{2} - \vec{Q} - \vec{q},s_{2}) \delta_{\vec{\underline{k}}_{1}+\vec{q},\vec{p}} \delta_{s_{1},s_{2}} \right] \nonumber \\
  & & - \frac{\sqrt{f_{c,s_{2}}(\vec{p})}}{A} \sum_{\vec{q}} U(\vec{q}) \times \nonumber \\
  & & \left\{ \frac{P_{\vec{Q}}(\vec{\underline{k}}_{1} - \vec{Q},s_{1})}{\sqrt{1-f_{c,s_{1}}(\vec{\underline{k}}_{1})}} \delta_{\vec{\underline{k}}_{2}-\vec{q},\vec{p}} \sqrt{1-f_{c,s_{2}}(\vec{\underline{k}}_{2})}  \right. \nonumber \\
  & & - \left. \frac{P_{\vec{Q}}(\vec{\underline{k}}_{2} - \vec{Q},s_{2})}{\sqrt{1-f_{c,s_{2}}(\vec{\underline{k}}_{2})}} \delta_{\vec{\underline{k}}_{1}+\vec{q},\vec{p}} \delta_{s_{1},s_{2}} \sqrt{1-f_{c,s_{1}}(\vec{\underline{k}}_{1})} \right\} \nonumber \\
  \label{eq:4bodyapp1}
\end{eqnarray}
The homogeneous part of the above equation is now a Hermitian eigenvalue equation. It has a complete set of orthonormal eigenfunctions $\phi^{tr}_{m,\vec{Q}}(\vec{k}_{1},s_{1};\vec{k}_{2},s_{2};\vec{p},s_{2})$. In the limit of very low electron density, when phase space filling effects can be ignored, and assuming $Q<<k_{F}$, the eigenenergies of the bound trion states can be expressed as,
\begin{equation}
  E^{tr}_{m}(\vec{Q},s_{1},s_{2}) = E^{tr}_{m}(\vec{Q}=0,s_{1},s_{2}) + \frac{\hbar^{2}Q^{2}}{2 m_{t}} 
\end{equation}
We should emphasize here that if the eigenstate in the form given in (\ref{eq:var}) is used as a variational state with the Hamiltonian given in (\ref{eq:H}) then the resulting coupled equations for the two-body and four-body correlations would be identical to the equations (\ref{eq:excitonapp1}) and (\ref{eq:4bodyapp1}) given in the Appendices.

\section{Appendix: Expression for $h_{n,u}(\vec{Q},\vec{p},\vec{q},s,s')$ in Equation (\ref{eq:exee})}  \label{app3}

The expression for $h_{n,u}(\vec{Q},\vec{p},\vec{q},s,s')$ appearing in (\ref{eq:exee}) is given below, 
\begin{eqnarray}
  & & \sqrt{2} h_{n,u}(\vec{Q},\vec{p},\vec{q},s,s') = \int \frac{d^{2}\vec{k}}{(2\pi)^{2}} V(\vec{q}) \sqrt{1 - f_{c,s'}(\vec{p} + \vec{q})} \nonumber \\
  & &  \times \sqrt{1 - f_{c,s}(\vec{k} + \lambda_{e} \vec{Q})} \sqrt{1 - f_{c,s}(\vec{k} + \lambda_{e} \vec{Q} - \vec{q})} \nonumber \\
  & & \times \phi^{ex*}_{n,\vec{Q}-\vec{q}}(\vec{k} - \lambda_{h}\vec{q}) \,\, \phi^{ex}_{n,\vec{Q}}(\vec{k}) \nonumber \\
  & & +u \int \frac{d^{2}\vec{k}}{(2\pi)^{2}} V(\vec{p} + \vec{q} - \lambda_{e}\vec{Q} -\vec{k}) \sqrt{1 - f_{c,s}(\vec{p} + \vec{q})} \nonumber \\
  & &  \times \sqrt{1 - f_{c,s}(\vec{k} + \lambda_{e} \vec{Q})} \sqrt{1 - f_{c,s'}(\vec{k} + \lambda_{e} \vec{Q} - \vec{q})} \nonumber \\
  & & \times \phi^{ex*}_{n,\vec{Q}-\vec{q}}(\vec{k} - \lambda_{h}\vec{q}) \,\, \phi^{ex}_{n,\vec{Q}}(\vec{k}) \nonumber \\
  & & - \int \frac{d^{2}\vec{k}}{(2\pi)^{2}} U(\vec{q}) \sqrt{1 - f_{c,s'}(\vec{p} + \vec{q})} \nonumber \\
  & &\times \phi^{ex*}_{n,\vec{Q}-\vec{q}}(\vec{k} + \lambda_{e}\vec{q}) \, \, \phi^{ex}_{n,\vec{Q}}(\vec{k}) \nonumber \\
  & & -u \int \frac{d^{2}\vec{k}}{(2\pi)^{2}} U(\vec{p} - \lambda_{e}\vec{Q} - \vec{k}) \sqrt{1 - f_{c,s'}(\vec{k} + \lambda_{e}\vec{Q})} \nonumber \\
  & & \times \phi^{ex*}_{n,\vec{Q}-\vec{q}}(\vec{k} + \lambda_{e}\vec{q}) \, \, \phi^{ex}_{n,\vec{Q}}(\vec{p} + \vec{q} - \lambda_{e} \vec{Q}) \nonumber \\
\end{eqnarray}
Here, $u = \pm 1$ depending on whether the exciton-electron scattering state (or the unbound trion state), as given in (\ref{eq:exelscat}), is symmetric or antisymmetric (consistent with the values of $s$ and $s'$).

\section{Appendix: The Importance of Including Exciton-Electron Scattering States (Unbound Trion States)} \label{app4}
An important point that needs to emphasized is that without including exciton-electron scattering, that takes spectral weight away from the higher energy peak as shown in Fig.~\ref{fig:fig5}, the peak optical conductivity of the lower energy peak can never exceed the peak conductivity of the higher energy peak. This follows from the basic physics of two coupled systems and can be seen as follows. Assuming $\gamma_{ex}=\gamma_{tr}=\gamma$ for simplicity, $\vec{Q}=0$, and ignoring exciton-electron scattering states, the poles of the exciton's Green's function will be at energies given by (\ref{eq:roots}) and are found to be,
\begin{equation}
\hbar \omega = \frac{E^{ex}_{0}+E^{tr}_{0}}{2} - i\gamma \pm \sqrt{(\frac{E^{ex}_{0}-E^{tr}_{0}}{2})^2 + |M_{0,0}|^2}  
\end{equation}
The corresponding spectral weights for the lower and the higher energy peaks in the optical absorption spectra would be proportional to $W$ and $1-W$, respectively, where,
\begin{equation}
  W = \frac{\displaystyle \sqrt{\left(\frac{E^{ex}_{0}-E^{tr}_{0}}{2}\right)^2 + |M_{0,0}|^2} - \left(\frac{E^{ex}_{0}-E^{tr}_{0}}{2}\right)}{\displaystyle 2\sqrt{\left(\frac{E^{ex}_{0}-E^{tr}_{0}}{2}\right)^2 + |M_{0,0}|^2}}
\end{equation}
When the electron density is zero, $M_{0,0}$ is zero, and the spectral weight all lies in the higher energy exciton peak in the optical absorption spectrum. As the electron density increases, the spectral weight begins to shift to the lower energy peak. But even when the electron density, and therefore $M_{0,0}$, are very large, the value of $W$ never exceeds $1/2$. Including the contribution of exciton-electron scattering states (or unbound trion states) is therefore necessary in producing the results shown in Fig.~\ref{fig:fig5} (where the peak optical conductivity of the lower energy peak is shown to become much larger than the peak optical conductivity of the higher energy peak at high electron densities) and accurately reproducing the experimental observations~\cite{Fai13,Chernikov15}.

\end{document}